\expandafter \def \csname CHAPLABELintro\endcsname {1}
\expandafter \def \csname CHAPLABELbackgrounds\endcsname {2}
\expandafter \def \csname CHAPLABELchains\endcsname {3}
\expandafter \def \csname EQLABELspectrum\endcsname {3.1?}
\expandafter \def \csname TABLABELBhiggs\endcsname {3.1?}
\expandafter \def \csname EQLABELCspectrum\endcsname {3.2?}
\expandafter \def \csname CHAPLABELpolyhedra\endcsname {4}
\expandafter \def \csname TABLABELscaling\endcsname {4.1?}
\expandafter \def \csname EQLABELWeier\endcsname {4.1?}
\expandafter \def \csname TABLABELSingularities\endcsname {4.2?}
\expandafter \def \csname FIGLABEL3tori\endcsname {4.1?}
\expandafter \def \csname TABLABELtops\endcsname {4.3?}
\expandafter \def \csname FIGLABEL3moretori\endcsname {4.2?}
\expandafter \def \csname TABLABELbots\endcsname {4.4?}
\expandafter \def \csname TABLABELABgps\endcsname {4.5?}
\expandafter \def \csname TABLABELChodgenos\endcsname {4.6?}
\expandafter \def \csname TABLABELallbref\endcsname {4.7?}
\expandafter \def \csname TABLABELBtm\endcsname {4.8?}
\expandafter \def \csname CHAPLABELfin\endcsname {5}
\expandafter \def \csname CHAPLABELappendix\endcsname {-2}
\expandafter \def \csname FIGLABELfigca1\endcsname {-2.1?}
\expandafter \def \csname FIGLABELfigca2\endcsname {-2.2?}
\expandafter \def \csname FIGLABELfigcca1\endcsname {-2.3?}
\expandafter \def \csname FIGLABELfigcca2\endcsname {-2.4?}


\font\eightrm=cmr8 at 8pt

\font\seventeenrm=cmr17 at 17pt
\font\twentyonerm=cmr17 at 21pt

\font\ss=cmss10

\font\csc=cmcsc10

\font\twelvecal=cmsy10 at 12pt

\font\twelvemath=cmmi12

\font\seventeenbold=cmbx7 at 17pt

\font\fively=lasy5
\font\sevenly=lasy7
\font\tenly=lasy10

\textfont10=\tenly
\scriptfont10=\sevenly
\scriptscriptfont10=\fively
\magnification=1200
\parskip=10pt
\parindent=20pt
\def\today{\ifcase\month\or January\or February\or March\or April\or May\or
June
       \or July\or August\or September\or October\or November\or December\fi
       \space\number\day, \number\year}

\def\title#1{\footline={\ifnum\pageno<2\hfil
       \else\hss\tenrm\folio\hss\fi}\vskip1truein\centerline{{#1}
       \footnote{\raise1ex\hbox{*}}{\eightrm Supported in part
       by the Robert A. Welch Foundation and N.S.F. Grants
       PHY-880637 and\break PHY-8605978.}}}

\def\newpage{\vfill\eject}
\def\abstract#1{\centerline{\bf ABSTRACT}\vskip.2truein{\narrower\noindent#1
       \smallskip}}

\def\runninghead#1#2{\voffset=2\baselineskip\nopagenumbers
       \headline={\ifodd\pageno\rightheadline\else \leftheadline\fi}
       \def\rightheadline{{\sl#1}\hfill{\rm\folio}}
       \def\leftheadline{{\rm\folio}\hfill{\sl#2}}}
\def\SS{\mathhexbox278}

\newcount\footnoteno
\def\Footnote#1{\advance\footnoteno by 1
                \let\SF=\empty
                \ifhmode\edef\SF{\spacefactor=\the\spacefactor}\/\fi
                $^{\the\footnoteno}$\ignorespaces
                \SF\vfootnote{$^{\the\footnoteno}$}{#1}}

\def\figbox#1#2#3{\vbox{\vskip15pt
                   \vbox{\hrule
                    \hbox{\vrule
                     \vbox{\vskip12truept\centerline #1 \vskip6truept
                          {\hskip.4truein\vbox{\hsize=5truein\noindent
                          {\bf Figure\hskip5truept#2:}\hskip5truept#3}}
                     \vskip18truept}
                    \vrule}
                   \hrule}}}
\def\place#1#2#3{\vbox to0pt{\kern-\parskip\kern-7pt
                             \kern-#2truein\hbox{\kern#1truein #3}
                             \vss}\nointerlineskip}
\def\figurecaption#1#2{\kern.75truein\vbox{\hsize=5truein\noindent{\bf Figure
    \figlabel{#1}:} #2}}
\def\tablecaption#1#2{\kern.75truein\lower12truept\hbox{\vbox{\hsize=5truein
    \noindent{\bf Table\hskip5truept\tablabel{#1}:} #2}}}
\def\boxed#1{\lower3pt\hbox{
                       \vbox{\hrule\hbox{\vrule

\vbox{\kern2pt\hbox{\kern3pt#1\kern3pt}\kern3pt}\vrule}
                         \hrule}}}

\def\g{\gamma}

\def\l{\lambda}
\def\m{\mu}
\def\n{\nu}

\def\ca#1{\relax\ifmmode {{\cal #1}}\else $\cal #1$\fi}

\def\calb{{\cal B}}

\def\calm{{\cal M}}

\def\inbar{\vrule height1.5ex width.4pt depth0pt}
\def\IB{\relax{\rm I\kern-.18em B}}
\def\IC{\relax\hbox{\kern.25em$\inbar\kern-.3em{\rm C}$}}
\def\ID{\relax{\rm I\kern-.18em D}}
\def\IE{\relax{\rm I\kern-.18em E}}
\def\IF{\relax{\rm I\kern-.18em F}}
\def\IG{\relax\hbox{\kern.25em$\inbar\kern-.3em{\rm G}$}}
\def\IH{\relax{\rm I\kern-.18em H}}
\def\II{\relax{\rm I\kern-.18em I}}
\def\IK{\relax{\rm I\kern-.18em K}}
\def\IL{\relax{\rm I\kern-.18em L}}
\def\IM{\relax{\rm I\kern-.18em M}}
\def\IN{\relax{\rm I\kern-.18em N}}
\def\IO{\relax\hbox{\kern.25em$\inbar\kern-.3em{\rm O}$}}
\def\IP{\relax{\rm I\kern-.18em P}}
\def\IQ{\relax\hbox{\kern.25em$\inbar\kern-.3em{\rm Q}$}}
\def\IR{\relax{\rm I\kern-.18em R}}
\def\IZ{\relax\ifmmode\hbox{\ss Z\kern-.4em Z}\else{\ss Z\kern-.4em Z}\fi}
\def\IGa{\relax{\rm I}\kern-.18em\Gamma}
\def\IPi{\relax{\rm I}\kern-.18em\Pi}
\def\ITh{\relax\hbox{\kern.25em$\inbar\kern-.3em\Theta$}}
\def\IOm{\relax\thinspace\inbar\kern1.95pt\inbar\kern-5.525pt\Omega}


\def\ie{{\it i.e.,\ \/}}

\def\noblackboxes{\overfullrule=0pt}

\def\cy{Calabi--Yau}
\def\cym{Calabi--Yau manifold}
\def\cys{Calabi--Yau manifolds}

\def\H#1#2{\relax\ifmmode {H^{#1#2}}\else $H^{#1 #2}$\fi}
\def\M{\relax\ifmmode{\calm}\else $\calm$\fi}

\def\Bigcheck{\lower3.8pt\hbox{\smash{\hbox{{\twentyonerm \v{}}}}}}
\def\bigboldcheck{\smash{\hbox{{\seventeenbold\v{}}}}}

\def\Bighat{\lower3.8pt\hbox{\smash{\hbox{{\twentyonerm \^{}}}}}}

\def\Msharp{\relax\ifmmode{\calm^\sharp}\else $\smash{\calm^\sharp}$\fi}
\def\Mflat{\relax\ifmmode{\calm^\flat}\else $\smash{\calm^\flat}$\fi}
\def\preMcheck{\kern2pt\hbox{\Bigcheck\kern-12pt{$\cal M$}}}
\def\Mcheck{\relax\ifmmode\preMcheck\else $\preMcheck$\fi}
\def\preMhat{\kern2pt\hbox{\Bighat\kern-12pt{$\cal M$}}}
\def\Mhat{\relax\ifmmode\preMhat\else $\preMhat$\fi}

\def\Bsharp{\relax\ifmmode{\calb^\sharp}\else $\calb^\sharp$\fi}
\def\Bflat{\relax\ifmmode{\calb^\flat}\else $\calb^\flat$ \fi}
\def\preBcheck{\hbox{\Bigcheck\kern-9pt{$\cal B$}}}
\def\Bcheck{\relax\ifmmode\preBcheck\else $\preBcheck$\fi}
\def\preBhat{\hbox{\Bighat\kern-9pt{$\cal B$}}}
\def\Bhat{\relax\ifmmode\preBhat\else $\preBhat$\fi}

\def\figBcheck{\kern3pt\hbox{\raise1pt\hbox{\bigboldcheck}\kern-11pt
    {\twelvecal B}}}
\def\figBsharp{{\twelvecal B}\raise5pt\hbox{$\twelvemath\sharp$}}
\def\figBflat{{\twelvecal B}\raise5pt\hbox{$\twelvemath\flat$}}

\def\gcheck{\hbox{\lower2.5pt\hbox{\Bigcheck}\kern-8pt$\g$}}
\def\lhat{\hbox{\raise.5pt\hbox{\Bighat}\kern-8pt$\l$}}

\def\Fcheck{\kern2pt\hbox{\raise1pt\hbox{\Bigcheck}\kern-10pt{$\cal F$}}}
\def\Fhat{\kern2pt\hbox{\raise1pt\hbox{\Bighat}\kern-10pt{$\cal F$}}}

\def\cp#1{\relax\ifmmode {\IP\kern-2pt{}_{#1}}\else $\IP\kern-2pt{}_{#1}$\fi}
\def\h#1#2{\relax\ifmmode {b_{#1#2}}\else $b_{#1#2}$\fi}

\def\frac#1#2{{#1\over #2}}

\def\cone{\relax\thinspace\hbox{$<\kern-.8em{)}$}}
\mathchardef\mho"0A30

\def\-{\hphantom{-}}



\def\picture #1 by #2 (#3){\vbox to #2{\hrule width #1 height 0pt depth 0pt
                                       \vfill\special{picture #3}}}
\def\scaledpicture #1 by #2 (#3 scaled #4){{\dimen0=#1 \dimen1=#2
           \divide\dimen0 by 1000 \multiply\dimen0 by #4
            \divide\dimen1 by 1000 \multiply\dimen1 by #4
            \picture \dimen0 by \dimen1 (#3 scaled #4)}}
\def\illustration #1 by #2 (#3){\vbox to #2{\hrule width #1 height 0pt depth
0pt
                                       \vfill\special{illustration #3}}}
\def\scaledillustration #1 by #2 (#3 scaled #4){{\dimen0=#1 \dimen1=#2
           \divide\dimen0 by 1000 \multiply\dimen0 by #4
            \divide\dimen1 by 1000 \multiply\dimen1 by #4
            \illustration \dimen0 by \dimen1 (#3 scaled #4)}}


\def\delaOssa{\nobreak\vskip1truein\hbox to\hsize
       {\hskip 4truein Xenia de la Ossa\hfill}}

\def\hoy{\number\day\space de \ifcase\month\or enero\or febrero\or marzo\or
       abril\or mayo\or junio\or julio\or agosto\or septiembre\or octubre\or
       noviembre\or diciembre\fi\space de \number\year}


\newif\ifproofmode
\proofmodefalse

\newif\ifforwardreference
\forwardreferencefalse

\newif\ifchapternumbers
\chapternumbersfalse

\newif\ifcontinuousnumbering
\continuousnumberingfalse

\newif\iffigurechapternumbers
\figurechapternumbersfalse

\newif\ifcontinuousfigurenumbering
\continuousfigurenumberingfalse

\newif\iftablechapternumbers
\tablechapternumbersfalse

\newif\ifcontinuoustablenumbering
\continuoustablenumberingfalse

\font\eqsixrm=cmr6

\def\marginstyle{\eqsixrm}

\newtoks\chapletter
\newcount\chapno
\newcount\eqlabelno
\newcount\figureno
\newcount\tableno

\chapno=0
\eqlabelno=0
\figureno=0
\tableno=0

\def\chapfolio{\ifnum\chapno>0 \the\chapno\else\the\chapletter\fi}

\def\bumpchapno{\ifnum\chapno>-1 \global\advance\chapno by 1
\else\global\advance\chapno by -1 \setletter\chapno\fi
\ifcontinuousnumbering\else\global\eqlabelno=0 \fi
\ifcontinuousfigurenumbering\else\global\figureno=0 \fi
\ifcontinuoustablenumbering\else\global\tableno=0 \fi}

\def\setletter#1{\ifcase-#1{}\or{}%
\or\global\chapletter={A}%
\or\global\chapletter={B}%
\or\global\chapletter={C}%
\or\global\chapletter={D}%
\or\global\chapletter={E}%
\or\global\chapletter={F}%
\or\global\chapletter={G}%
\or\global\chapletter={H}%
\or\global\chapletter={I}%
\or\global\chapletter={J}%
\or\global\chapletter={K}%
\or\global\chapletter={L}%
\or\global\chapletter={M}%
\or\global\chapletter={N}%
\or\global\chapletter={O}%
\or\global\chapletter={P}%
\or\global\chapletter={Q}%
\or\global\chapletter={R}%
\or\global\chapletter={S}%
\or\global\chapletter={T}%
\or\global\chapletter={U}%
\or\global\chapletter={V}%
\or\global\chapletter={W}%
\or\global\chapletter={X}%
\or\global\chapletter={Y}%
\or\global\chapletter={Z}\fi}

\def\tempsetletter#1{\ifcase-#1{}\or{}%
\or\global\chapletter={A}%
\or\global\chapletter={B}%
\or\global\chapletter={C}%
\or\global\chapletter={D}%
\or\global\chapletter={E}%
\or\global\chapletter={F}%
\or\global\chapletter={G}%
\or\global\chapletter={H}%
\or\global\chapletter={I}%
\or\global\chapletter={J}%
\or\global\chapletter={K}%
\or\global\chapletter={L}%
\or\global\chapletter={M}%
\or\global\chapletter={N}%
\or\global\chapletter={O}%
\or\global\chapletter={P}%
\or\global\chapletter={Q}%
\or\global\chapletter={R}%
\or\global\chapletter={S}%
\or\global\chapletter={T}%
\or\global\chapletter={U}%
\or\global\chapletter={V}%
\or\global\chapletter={W}%
\or\global\chapletter={X}%
\or\global\chapletter={Y}%
\or\global\chapletter={Z}\fi}

\def\chapshow#1{\ifnum#1>0 \relax#1%
\else{\tempsetletter{\number#1}\chapno=#1\chapfolio}\fi}

\def\chaplabel#1{\bumpchapno\ifproofmode\ifforwardreference
\immediate\write1{\noexpand\expandafter\noexpand\def
\noexpand\csname CHAPLABEL#1\endcsname{\the\chapno}}\fi\fi
\global\expandafter\edef\csname CHAPLABEL#1\endcsname
{\the\chapno}\ifproofmode\llap{\hbox{\marginstyle #1\ }}\fi\chapfolio}

\def\eqnum{\global\advance\eqlabelno by 1
\eqno(\ifchapternumbers\chapfolio.\fi\the\eqlabelno)}

\def\eqlabel#1{\global\advance\eqlabelno by 1 \ifproofmode\ifforwardreference
\immediate\write1{\noexpand\expandafter\noexpand\def
\noexpand\csname EQLABEL#1\endcsname{\the\chapno.\the\eqlabelno?}}\fi\fi
\global\expandafter\edef\csname EQLABEL#1\endcsname
{\the\chapno.\the\eqlabelno?}\eqno(\ifchapternumbers\chapfolio.\fi
\the\eqlabelno)\ifproofmode\rlap{\hbox{\marginstyle #1}}\fi}

\def\eqalignnum{\global\advance\eqlabelno by 1
&(\ifchapternumbers\chapfolio.\fi\the\eqlabelno)}

\def\eqalignlabel#1{\global\advance\eqlabelno by 1 \ifproofmode
\ifforwardreference\immediate\write1{\noexpand\expandafter\noexpand\def
\noexpand\csname EQLABEL#1\endcsname{\the\chapno.\the\eqlabelno?}}\fi\fi
\global\expandafter\edef\csname EQLABEL#1\endcsname
{\the\chapno.\the\eqlabelno?}&(\ifchapternumbers\chapfolio.\fi
\the\eqlabelno)\ifproofmode\rlap{\hbox{\marginstyle #1}}\fi}

\def\eqref#1{\hbox{(\ifundefined{EQLABEL#1}***)\ifproofmode\ifforwardreference%
\else\write16{ ***Undefined Equation Reference #1*** }\fi
\else\write16{ ***Undefined Equation Reference #1*** }\fi
\else\edef\LABxx{\getlabel{EQLABEL#1}}%
\def\LAByy{\expandafter\stripchap\LABxx}\ifchapternumbers%
\chapshow{\LAByy}.\expandafter\stripeq\LABxx%
\else\ifnum\number\LAByy=\chapno\relax\expandafter\stripeq\LABxx%
\else\chapshow{\LAByy}.\expandafter\stripeq\LABxx\fi\fi)\fi}%
\ifproofmode\write2{Equation #1}\fi}

\def\fignum{\global\advance\figureno by 1
\relax\iffigurechapternumbers\chapfolio.\fi\the\figureno}

\def\figlabel#1{\global\advance\figureno by 1
\relax\ifproofmode\ifforwardreference
\immediate\write1{\noexpand\expandafter\noexpand\def
\noexpand\csname FIGLABEL#1\endcsname{\the\chapno.\the\figureno?}}\fi\fi
\global\expandafter\edef\csname FIGLABEL#1\endcsname
{\the\chapno.\the\figureno?}\iffigurechapternumbers\chapfolio.\fi
\ifproofmode\llap{\hbox{\marginstyle#1
\kern1.2truein}}\relax\fi\the\figureno}

\def\figref#1{\hbox{\ifundefined{FIGLABEL#1}!!!!\ifproofmode\ifforwardreference%
\else\write16{ ***Undefined Figure Reference #1*** }\fi
\else\write16{ ***Undefined Figure Reference #1*** }\fi
\else\edef\LABxx{\getlabel{FIGLABEL#1}}%
\def\LAByy{\expandafter\stripchap\LABxx}\iffigurechapternumbers%
\chapshow{\LAByy}.\expandafter\stripeq\LABxx%
\else\ifnum \number\LAByy=\chapno\relax\expandafter\stripeq\LABxx%
\else\chapshow{\LAByy}.\expandafter\stripeq\LABxx\fi\fi\fi}%
\ifproofmode\write2{Figure #1}\fi}

\def\tabnum{\global\advance\tableno by 1
\relax\iftablechapternumbers\chapfolio.\fi\the\tableno}

\def\tablabel#1{\global\advance\tableno by 1
\relax\ifproofmode\ifforwardreference
\immediate\write1{\noexpand\expandafter\noexpand\def
\noexpand\csname TABLABEL#1\endcsname{\the\chapno.\the\tableno?}}\fi\fi
\global\expandafter\edef\csname TABLABEL#1\endcsname
{\the\chapno.\the\tableno?}\iftablechapternumbers\chapfolio.\fi
\ifproofmode\llap{\hbox{\marginstyle#1
\kern1.2truein}}\relax\fi\the\tableno}

\def\tabref#1{\hbox{\ifundefined{TABLABEL#1}!!!!\ifproofmode\ifforwardreference%
\else\write16{ ***Undefined Table Reference #1*** }\fi
\else\write16{ ***Undefined Table Reference #1*** }\fi
\else\edef\LABtt{\getlabel{TABLABEL#1}}%
\def\LABTT{\expandafter\stripchap\LABtt}\iftablechapternumbers%
\chapshow{\LABTT}.\expandafter\stripeq\LABtt%
\else\ifnum\number\LABTT=\chapno\relax\expandafter\stripeq\LABtt%
\else\chapshow{\LABTT}.\expandafter\stripeq\LABtt\fi\fi\fi}%
\ifproofmode\write2{Table#1}\fi}

\newdimen\sectionskip     \sectionskip=20truept
\newcount\sectno
\def\section#1#2{\sectno=0 \null\vskip\sectionskip
    \centerline{\chaplabel{#1}.~~{\bf#2}}\nobreak\vskip.2truein
    \noindent\ignorespaces}

\def\advancesectno{\global\advance\sectno by 1}
\def\sectfolio{\number\sectno}
\def\subsection#1{\goodbreak\advancesectno\null\vskip10pt
                  \noindent\chapfolio.~\sectfolio.~{\bf #1}
                  \nobreak\vskip.05truein\noindent\ignorespaces}

\def\uttg#1{\null\vskip.1truein
    \ifproofmode \line{\hfill{\bf Draft}:
    UTTG--{#1}--\number\year}\line{\hfill\today}
    \else \line{\hfill UTTG--{#1}--\number\year}
    \line{\hfill\ifcase\month\or January\or February\or March\or April\or
May\or June
    \or July\or August\or September\or October\or November\or December\fi
    \space\number\year}\fi}

\def\getlabel#1{\csname#1\endcsname}
\def\ifundefined#1{\expandafter\ifx\csname#1\endcsname\relax}
\def\stripchap#1.#2?{#1}
\def\stripeq#1.#2?{#2}

%
\catcode`@=11 
\def\space@ver#1{\let\@sf=\empty\ifmmode#1\else\ifhmode%
\edef\@sf{\spacefactor=\the\spacefactor}\unskip${}#1$\relax\fi\fi}
\newcount\referencecount     \referencecount=0
\newif\ifreferenceopen       \newwrite\referencewrite
\newtoks\rw@toks
\def\refmark#1{\relax[#1]}
\def\refend{\refmark{\number\referencecount}}
\newcount\lastrefsbegincount \lastrefsbegincount=0
\def\refsend{\refmark{\count255=\referencecount%
\advance\count255 by -\lastrefsbegincount%
\ifcase\count255 \number\referencecount%
\or\number\lastrefsbegincount,\number\referencecount%
\else\number\lastrefsbegincount-\number\referencecount\fi}}
\def\refch@ck{\chardef\rw@write=\referencewrite
\ifreferenceopen\else\referenceopentrue
\immediate\openout\referencewrite=referenc.texauxil \fi}
%
{\catcode`\^^M=\active 
  \gdef\obeyendofline{\catcode`\^^M\active \let^^M\ }}%
%
{\catcode`\^^M=\active 
  \gdef\ignoreendofline{\catcode`\^^M=5}}
{\obeyendofline\gdef\rw@start#1{\def\t@st{#1}\ifx\t@st\blankend%
\endgroup\@sf\relax\else\ifx\t@st\bl@nkend\endgroup\@sf\relax%
\else\rw@begin#1
\backtotext
\fi\fi}}
{\obeyendofline\gdef\rw@begin#1
{\def\n@xt{#1}\rw@toks={#1}\relax%
\rw@next}}
\def\blankend{}
{\obeylines\gdef\bl@nkend{
}}
\newif\iffirstrefline  \firstreflinetrue
\def\rwr@teswitch{\ifx\n@xt\blankend\let\n@xt=\rw@begin%
\else\iffirstrefline\global\firstreflinefalse%
\immediate\write\rw@write{\noexpand\obeyendofline\the\rw@toks}%
\let\n@xt=\rw@begin%
\else\ifx\n@xt\rw@@d \def\n@xt{\immediate\write\rw@write{%
\noexpand\ignoreendofline}\endgroup\@sf}%
\else\immediate\write\rw@write{\the\rw@toks}%
\let\n@xt=\rw@begin\fi\fi\fi}
\def\rw@next{\rwr@teswitch\n@xt}
\def\rw@@d{\backtotext} \let\rw@end=\relax
\let\backtotext=\relax

\newdimen\refindent     \refindent=30pt
\def\Textindent#1{\noindent\llap{#1\enspace}\ignorespaces}
\def\refitem#1{\par\hangafter=0 \hangindent=\refindent\Textindent{#1}}
\def\REFNUM#1{\space@ver{}\refch@ck\firstreflinetrue%
\global\advance\referencecount by 1 \xdef#1{\the\referencecount}}
\def\refnum#1{\space@ver{}\refch@ck\firstreflinetrue%
\global\advance\referencecount by 1\xdef#1{\the\referencecount}\refend}

\def\REF#1{\REFNUM#1%
\immediate\write\referencewrite{%
\noexpand\refitem{#1.}}%
\begingroup\obeyendofline\rw@start}
\def\ref{\refnum\?%
\immediate\write\referencewrite{\noexpand\refitem{\?.}}%
\begingroup\obeyendofline\rw@start}
\def\Ref#1{\refnum#1%
\immediate\write\referencewrite{\noexpand\refitem{#1.}}%
\begingroup\obeyendofline\rw@start}
\def\REFS#1{\REFNUM#1\global\lastrefsbegincount=\referencecount%
\immediate\write\referencewrite{\noexpand\refitem{#1.}}%
\begingroup\obeyendofline\rw@start}

\def\REFSCON#1{\REF#1}

\def\cite#1{\refmark#1}
\catcode`@=12 
%
\input epsf.tex
\proofmodefalse
\baselineskip=15pt plus 1pt minus 1pt
\parskip=5pt
\chapternumberstrue
\figurechapternumberstrue
\tablechapternumberstrue
\ifproofmode
\immediate\openout2=allcrossreferfile \fi
\ifforwardreference\input labelfile
\ifproofmode\immediate\openout1=labelfile \fi\fi
\noblackboxes
\hfuzz=1pt
\vfuzz=2pt


\def\hourandminute{\count255=\time\divide\count255 by 60
\xdef\hour{\number\count255}
\multiply\count255 by -60\advance\count255 by\time
\hour:\ifnum\count255<10 0\fi\the\count255}

\def\subsection#1{\goodbreak\advancesectno\null\vskip10pt
                  \noindent{\it \chapfolio.\sectfolio.~#1}
                  \nobreak\vskip.05truein\noindent\ignorespaces}
\def\cite#1{\refmark{#1}}
\def\\{\hfill\break}

\def\point#1{\noindent\setbox0=\hbox{#1}\kern-\wd0\box0}

\nopagenumbers\pageno=0
\rightline{\eightrm UTTG-11-97}\vskip-5pt
\rightline{\eightrm hep-th/9703148}\vskip-5pt
\rightline{\eightrm  March 20 1997}

\vskip1truein
\centerline{\seventeenrm Comments on A,B,C Chains}
\vskip10pt
\centerline{\seventeenrm of Heterotic and Type II Vacua}
\vskip0.5truein
\centerline{\csc Philip~Candelas$^1$, Eugene~Perevalov$^2$ and
             Govindan~Rajesh$^3$}
\vfootnote{$^{\eightrm 1}$}{\eightrm candelas@physics.utexas.edu.}
\vfootnote{$^{\eightrm 2}$}{\eightrm pereval@physics.utexas.edu.}
\vfootnote{$^{\eightrm 3}$}{\eightrm rajesh@physics.utexas.edu.}

\bigskip
\centerline{\it Theory Group}
\centerline{\it Department of Physics}
\centerline{\it University of Texas}
\centerline{\it Austin, TX 78712, USA}
\vskip1in\bigskip
\nobreak\vbox{
\centerline{\bf ABSTRACT}
\vskip.25truein
\noindent{ We construct, as hypersurfaces in toric varieties, \cys\
corresponding to F-theory vacua dual to  $E_8\times E_8$ heterotic strings
compactified to six dimensions on $K3$ surfaces with non-semisimple gauge
backgrounds. These vacua were studied in the recent work of Aldazabal, Font,
Ib\'{a}\~{n}ez and Uranga. We extend their results by constructing many more
examples, corresponding to enhanced gauge symmetries, by noting that they can
be obtained from previously known \cys\ corresponding to $K3$ compactification
of heterotic strings with simple gauge backgrounds by means of extremal
transitions of the conifold type.
}
}
\newpage
    {\bf Contents}
\vskip5pt

    1. Introduction
\vskip3pt

    2. Simple and Non-Semisimple Gauge Backgrounds on $K3$
\vskip3pt

    3. Construction of the B and C Chains on the Heterotic Side
\vskip3pt

    4. \cys\ corresponding to the B and C Chains
\vskip3pt

\hskip10pt 4.1 {\it Constructing enhanced gauge groups for the A series}

\hskip10pt 4.2 {\it \cy\ Manifolds for the B and C series}

\hskip10pt 4.3 {\it Conifold transitions}

\hskip10pt 4.4 {\it Nonperturbative vacua --- tensor multiplets}
\vskip3pt

    5. Discussion
\vskip3pt

    Appendix: Figures
\newpage
\pageno=1
\headline={\ifproofmode\hfil\eightrm draft:\ \today\
\hourandminute\else\hfil\fi}
\footline={\rm\hfil\folio\hfil}
\section{intro}{Introduction}
F-theory ~\REFS\rVH{C.~Vafa, Nucl.Phys. B469 (1996) 403-418, hep-th/9602022; \\
C.~Hull, Nucl. Phys. B468 (1996) 113, hep-th/9512181.}
\refsend\
has proved to be a powerful tool for analysing string dualities. In
particular it provides the basis for a geometric understanding of string
dualities. For
example, it was argued in~\REFS\rMV{D.~R.~Morrison and C.~Vafa,
Nucl.Phys. B473 (1996) 74-92, hep-th/9602114;\\
D.~R.~Morrison and C.~Vafa,
Nucl.Phys. B476 (1996) 437, hep-th/9603161.} \refsend\
that F-theory compactified on an elliptic \cym\ is dual to the heterotic
$E_8\times E_8$ theory on a $K3$ surface. Furthermore it was shown that the
base of the elliptic \cym\ is a Hirzebruch surface $\IF_{n}$~, or a blowup
thereof. Upon further toroidal compactification to four dimensions, we obtain
Type II/heterotic duality which has been well studied recently
{}~\REFS\rMany{S.~Kachru and C.~Vafa, Nucl. Phys. B450 (1995) 69, hep-th/
9505105;\\
S. Ferrara, J. Harvey, A. Strominger and C.~Vafa,\\Phys. Lett. B361
(1995) 59 hep-th/9505162;\\
G.~Aldazabal, A.~Font, L.~E.~Ib\'{a}\~{n}ez and F.~Quevedo,\\ Nucl. Phys. B461
(1996) 85, hep-th/9510093; \\
A.~Klemm, W.~Lerche and P.~Mayr, Phys. Lett. B357 (1995) 313, hep-th/9506112;
\\
P.~S.~Aspinwall and J.~Louis, Phys. Lett. B369 (1996) 233, hep-th/9510234;\\
P.~S.~Aspinwall, Phys. Lett. B371 (1996) 231, hep-th/9511171;\\
P.~S.~Aspinwall and M.~Gross, Phys.Lett. B382 (1996) 81-88, hep-th/9602118;\\
C.~Vafa and E.~Witten, hep-th/9507050;\\
C.~G\'{o}mez and E.~L\'{o}pez, Phys. Lett. B356 (1995) 487,
hep-th/9506024;\\
M.~Bill\'{o} et. al., Class. Quant. Grav. 13 (1996) 831, hep-th/9506075;\\
I.~Antoniadis, E.~Gava, K.~S.~Narain and K.~R.~Taylor,\\ Nucl. Phys. B455
(1995) 109, hep-th/9507155;\\
G.~Lopes Cardoso, D.~L\"{u}st and T.~Mohaupt,\\ Nucl. Phys. B455 (1995) 131,
hep-th/9507113;\\
G.~Curio, Phys. Lett. B368 (1996) 78, hep-th/9509146;\\
V.~Kaplunovsky, J.~Louis and S.~Theisen,\\ Phys. Lett. 357B (1995) 71,
hep-th/9506110;\\
S.~Kachru, A.~Klemm, W.~Lerche, P.~Mayr and C.~Vafa,\\ Nucl. Phys B459 (1996)
537, hep-th/9508155;\\
B.~Hunt and R.~Schimmrigk, hep-th/9512138;\\
R.~Blumenhagen and A.~Wisskirchen, hep-th/9601050;\\
E.~G.~Gimon and C.~V.~Johnson, Nucl.Phys. B479 (1996) 285-304, hep-th/9606176.}
\refsend .
Many examples of \cys\ corresponding to six dimensional
F-theory/heterotic dual pairs were obtained as hypersurfaces in toric
varieties in~
\REFS\rCF{P.~Candelas and A.~Font, hep-th/9603170.}
\refsend , where the gauge group in the effective theory appeared as a result
of embedding simple gauge backgrounds in each $E_8$ factor on the heterotic
side. These will be referred to as A models.
These examples were shown to be organized in the form of chains obtained
on the heterotic side by sequential Higgsing of the commutant of the gauge
background in $E_8\times E_8$. Furthermore, the polyhedra encoding the toric
data of these manifolds were shown to exhibit a regular structure related in a
simple way to the unbroken gauge symmetry of their heterotic duals.

In a recent work, Aldazabal {\it et al.}~
\REFS\rABCD{G.~Aldazabal, A.~Font, L.~E.~Ib\'{a}\~{n}ez and
A.~M.~Uranga, hep-th/9607121.}\refsend\ discuss an interesting class of models,
which they call the B, C and D series, in which the gauge
backgrounds are non-semisimple (\ie the backgrounds included one or more
$U(1)$ factors). However, only a few members of each chain in this
class were obtained. In this paper, we realise a large
number of members of these chains as hypersurface in toric varieties by
relating them to previously known models
by extremal transitions of the conifold type. The fact that the models are
related by extremal transitions was also noted in \cite\rABCD \ for their
original examples, though it was not used to construct them. The point
emphasized here, as in~\cite\rCF, is that toric geometry is the natural arena
in which to discuss these webs of vacua.

This paper is organized as follows. In \SS{2}, we give a brief
review of
simple and nonsemisimple gauge backgrounds on $K3$, followed in \SS{3} by a
review of the construction of the A, B and C chains on the heterotic side.
In \SS{4}, we demonstrate our method for constructing \cys\ corresponding
to the B and C chains, and compare the Hodge numbers of various members
obtained using the methods of Batyrev~
\REFS\rBat{V.~Batyrev, Duke Math. Journ. 69 (1993) 349.}
\REFSCON\rCOK{P.~Candelas, X.~de la Ossa and S.~Katz,\\ Nucl. Phys. B450 (1995)
267, hep-th/9412117.}
\refsend\
with the heterotic results. We also construct \cys\ corresponding to models
with extra tensor multiplets. \SS{5} concludes with a brief discussion of
our results, and an Appendix contains figures of the polyhedra that we discuss.

\newpage

\section{backgrounds}{Simple and Non-Semisimple Backgrounds on $K3$}
In this section we give a brief review, following~\cite\rABCD , of the
construction of heterotic models
in six and four dimensions by compactifying $E_8\times E_8$ heterotic string
on $K3$ and $K3\times T^2$, respectively.
First, consider the case of simple
gauge backgrounds. Let $H_{1,2}$ be the background gauge groups (simple
subgroups of $E_8$) and $k_{1,2}$ the corresponding instanton numbers (second
Chern classes of the background gauge bundles on the $K3$). The contribution
of each $E_8$ to the unbroken gauge group in six dimensions is then the
commutant $G_{1,2}$ of $H_{1,2}$ respectively. The number of hypermultiplets
in the representation $R_a$ of $G$ is then
$$N(R_a)=kT(M_a)-{\rm dim}(M_a),$$
where the adjoint of $E_8$ decomposes under $G\times H$ as {\bf 248}=$\sum_{a}
(R_a, M_a)$, and $T(M_a)$ is given by
tr$(T_{a}^{i}T_{a}^{j})=T(M_a)\delta _{ij}$,
$T_{a}^{i}$ being a generator of $H$ in the representation $M_a$.

Anomaly cancellation requires $k_1+k_2=24$, so it is convenient to define
$$n=k_1-12=12-k_2$$ and take $n\ge 0$ (\ie $k_1\ge k_2$). If $n\le 8$ we can
take $H_1=H_2=SU(2)$ and
obtain $E_7\times E_7$ gauge symmetry in six dimensions
with the following matter content:
$${1\over 2}(8+n)({\bf 56}, {\bf 1})+{1\over 2}(8-n)({\bf 1}, {\bf 56})
+ 62({\bf 1}, {\bf 1})$$
If $9\le n\le 12$, $k_2$ cannot support an $SU(2)$ background, and the
instantons in the second $E_8$ are necessarily small producing an unbroken
$E_8$. The gauge group in six dimensions is thus $E_7\times E_8$ with matter
content
$${1\over 2}(8+n)({\bf 56}, {\bf 1})+(53+n)({\bf 1}, {\bf 1})$$
Models with subgroups of the above can be obtained by gauge symmetry
breaking via Higgs mechanism, or, equivalently, by taking the subgroups of
$E_8$ other than $SU(2)$ as~$H_{1,2}$.

There exists another possibility for constructing heterotic models in six
dimensions which was first considered in ref.~
\REFS\rGSW{M. B. Green, J. H. Schwarz and P. C. West,
Nucl. Phys. B254 (1985) 327.}
\refsend
{}.
It consists of considering $U(1)$ (\ie nonsemisimple) backgrounds in each
$E_8$ and proceeds as follows. The instanton number of the $U(1)$
configuration is taken to be
$$m_i={1\over 16\pi ^2}\int_{K3} {1\over 30}{\rm Tr}F_{U(1)_i}^2~,~~~ i=1,2$$
and anomaly cancellation again requires $m_1+m_2=24$.

Now the adjoint of $E_8$ decomposes under $E_7\times U(1)$ as
$${\bf 248}=({\bf 133}, 0)+({\bf 56}, q)+({\bf 56}, -q)+({\bf 1}, 2q)+
({\bf 1}, -2q)+({\bf 1}, 0)$$
The generator $Q$ of $U(1)$ is normalized as a generator of $E_8$ in the
adjoint,
so $q={1\over 2}$. The index theorem applied to this case gives
$$N_q=mq^2 - 1,$$
for the number, $N_q$, of hypermultiplets of charge $q$.

Thus the matter content of the resulting $E_7\times U(1)\times E_7\times U(1)$
turns out to be
$$\eqalign{ & \{ {1\over 4}(m_1-4)({\bf 56},{1\over 2};{\bf 1},0)+{1\over 4}
           (m_2-4)({\bf 1},0;{\bf 56},{1\over 2})+ \cr
         &\qquad (m_1-1)({\bf 1},1;{\bf 1},0)+(m_2-1)({\bf 1},0;{\bf 1},1)+
           {\rm c.c.} \}
          +20({\bf 1},0;{\bf 1},0)\cr }$$
It was further noticed in~\cite\rABCD ~that these $U(1)$'s are such that the
anomaly 8-form does not in general factorize into a product of two 4-forms and
hence the residual anomaly cannot be cancelled by the Green-Schwarz mechanism.
The anomaly can, however, be cancelled completely for a certain linear
combination
of the above $U(1)$'s. The orthogonal combination is still anomalous, and must
be spontaneously broken if we want to obtain a consistent low energy theory.
The mechanism for such breaking was described in refs.~\cite\rGSW ~and~
\REFS\rEW{E. Witten, Phys. Lett. B149 (1984) 351.}\refsend. One linear
combination of the two photons becomes massive by swallowing
a $B$-field zero mode. Moreover, in our case, the combination which gets mass
is the anomalous one. So the actual gauge group is $E_7\times E_7\times U(1)$,
and the anomaly is absent.

We can now combine both abelian and non-abelian backgrounds and obtain
$H\times
U(1)$ bundles with instanton numbers $(k, m)$ in each $E_8$ . The gauge group
is then $G\times U(1)$, with $G$ the commutant of $H\times U(1)$ in $E_8$,
and the adjoint of $E_8$ decomposes as $ {\bf 248} = \sum _a (R_a, q_a, M_a)$
under $G\times U(1)\times H$. The number of hypermultiplets in
the
$(R_a, q_a)$ representation of $G\times U(1)$ is again given by the index
theorem $$ N(R_a, q_a) = kT(M_a) + (mq_a^2 -1) \ {\rm dim} \ M_a$$
where we again normalize Tr$Q^2 = 30$. This may now be generalized in a
straightforward way to the case of $H\times U(1)^{8-d}$ bundles with
$d\le 6$.
In the following section, we consider two choices leading to the B and C type
models.

\section{chains}{Construction of the B and C Chains on the Heterotic Side}
This section summarises the construction of the B and C chains on the heterotic
side as described in \cite\rABCD.

The B type chains can be constructed by embedding $(3,\, 3)$ $U(1)$ instantons
and
\hbox{$(9+n,\, 9-n)$} $SU(2)$ instantons in the two $E_8$'s. The commutant in
each
$E_8$ of this background is $E_6\times U(1)$. The unbroken $U(1)$ gauge group
is then the diagonal combination, $U(1)_D$, of the two $U(1)$'s.

The $E_6\times E_6 \times U(1)_D$ spectrum is given in \cite\rABCD ~as:
$$\eqalign
{& \{  {1\over 2} (k_1 - 3)({\bf 27},{\bf 1},{1\over {2\sqrt 6} }) +
    {1\over 2} (k_2 - 3)({\bf 1},{\bf 27},{1\over {2\sqrt 6} }) + \cr
 &  \qquad {1\over 2} (k_1 + k_2 +10)
({\bf 1},{\bf 1},{3 \over {2\sqrt 6}}) +
   {\rm c.c.}  \}  + (2k_1 + 2k_2 + 13)({\bf 1},{\bf 1},0) \cr}
\eqlabel{spectrum}$$
where $ k_1 = 9 + n$ and $k_2 = 9 - n$.
The Hodge numbers of the \cym\ for the F-theory dual of this model would
be$^{\ast}$ $(h_{21}, h_{11}) = (48, 16)$.
Given this spectrum, we can systematically Higgs the $E_6$ gauge groups as
mentioned in \cite\rABCD ~and~
\Ref\rLetter{A.~Font, private communication.}.
Upon Higgsing the first $E_6$, the
Hodge numbers change as shown in the first part of Table~\tabref{Bhiggs}.
\footnote\ {$^{\ast}$These are correct only for $n\le 6$. For $n>6$, the group
is no longer $E_6\times E_6 \times U(1)$, but $E_6\times E_7\times U(1)$. For
$n=7$, the Hodge numbers are (48, 18), the extra vector multiplet in the 4D
Coulomb branch coming from a small instanton. For $n=8$, there is no small
instanton, and the Hodge numbers are (49, 17). }

The C type chains are constructed by embedding $SU(2)\times
U(1)_1\times U(1)_2$ instantons in each $E_8$, with instanton numbers
$(7+n,3,2;7-n,3,2)$. The commutant in each $E_8$ of this background is
$SO(10)\times U(1)^2$. The unbroken $U(1)\times U(1)$ gauge group
consists of the diagonal combinations of the $U(1)^2$ groups from
each $E_8$.

\newpage
$${
\vbox{\offinterlineskip\halign{
\strut # height 10pt depth 5pt&\quad$#$\quad\hfil\vrule
&\quad$#$\quad\hfil\vrule\cr
\noalign{\hrule}
\vrule&\hfil $Group$ &\hfil (\Delta h_{21},\, \Delta h_{11})\cr
\noalign{\hrule\vskip3pt\hrule}
\vrule& SO(10)& (k_1 - 4,\, -1)\cr
\vrule& SU(5) & (k_1 - 6,\, -1)\cr
\vrule& SU(4) & (2k_1 - 9,\, -1)\cr
\vrule& SU(3) & (2k_1 - 11,\, -1)\cr
\vrule& SU(2) & (4k_1 - 25,\, -1)\cr
\vrule& SU(1) & (6k_1 - 39,\, -1)\cr
\noalign{\hrule}
}}
\hskip20pt
\vbox{\offinterlineskip\halign{
\strut # height 10pt depth 5pt&\quad$#$\quad\hfil\vrule
&\quad$#$\quad\hfil\vrule\cr
\noalign{\hrule}
\vrule&\hfil $Group$ &\hfil (\Delta h_{21},\, \Delta h_{11})\cr
\noalign{\hrule\vskip3pt\hrule}
\vrule& & \cr
\vrule& SU(5) & (k_1 - 4,\, -1)\cr
\vrule& SU(4) & (k_1 - 2,\, -1)\cr
\vrule& SU(3) & (2k_1 - 7,\, -1)\cr
\vrule& SU(2) & (2k_1 - 9,\, -1)\cr
\vrule& SU(1) & (4k_1 - 17,\, -1)\cr
\noalign{\hrule}
}}
}$$
\nobreak\tablecaption{Bhiggs}{The change in the Hodge numbers of the B and C
models
obtained upon sequential Higgsing of the first $E_6$($SO(10)$ for the C
models) to subgroups. The entry
in each row gives the change in the Hodge numbers from the Hodge numbers
corresponding to the group in the previous row. Here $k_1$ is the number of
$SU(2)$ instantons in the first $E_8$. Higgsing is possible only when
$\Delta h_{21}$ is nonnegative. For the second $E_6$ (or $SO(10)$), replace
replace $k_1$ by $k_2$.}
\bigskip

The $SO(10)\times SO(10)\times U(1)\times U(1)$ spectrum is given in
\cite\rABCD ~as:
$$\eqalign
{&\{ {1\over 2} (k_1 - 3)({\bf 16},{\bf 1},0,-{1\over 4} ) +
{1\over 2} (k_2 - 3)({\bf 1},{\bf 16},0,-{1\over 4} ) \cr
&~~+ {1\over 2} (k_1 - 1)({\bf 10},{\bf 1},{1\over {2\sqrt 2}},0 ) +
{1\over 2} (k_2 - 1)({\bf 1},{\bf 10},{1\over {2\sqrt 2}},0 )  \cr
 &~~~~+ {1\over 2} (k_1 + k_2 + 3)
   [({\bf 1},{\bf 1},{1\over {2\sqrt 2}},-{1\over 2}) +
    ({\bf 1},{\bf 1},{1\over {2\sqrt 2}},{1\over 2})] +
   4({\bf 1},{\bf 1},{1\over {2\sqrt 2}},0) + {\rm c.c.} \} \cr
 &~~~~~~+ (2k_1 + 2k_2 + 12)({\bf 1},{\bf 1},0,0) \cr}
\eqlabel{Cspectrum}$$
with $ k_1 = 7 + n$ and $k_2 = 7 - n$.
The Hodge numbers of the \cym\ for the F-theory dual of this model would
be$^{\ast}$ $(h_{21}, h_{11}) = (39, 15)~$\cite\rABCD. Upon Higgsing the first
$SO(10)$, the Hodge numbers change as shown in the second part of
Table~\tabref{Bhiggs}.
\footnote\ {$^{\ast}$These are correct only for $n\le 4$. For $n>4$, the group
is no longer $SO(10)\times SO(10) \times U(1)^2$, but
$SO(10)\times E_6\times U(1)^2$. For $n=5$, the Hodge numbers are (39, 17),
the extra vector multiplet in the 4D Coulomb branch coming from a small
instanton. For $n=6$, there is no small instanton, and the Hodge numbers are
(40, 16).}

\newpage
\section{polyhedra}{\cy\ Manifolds corresponding to the B and C Chains}
\vskip-20pt
\subsection{Constructing enhanced gauge groups}
We give a brief review of the procedures used to construct \cys\
corresponding to enhanced gauge symmetries in the A chains.
The starting point is the construction of the lowest members of these chains
as elliptic fibrations over the Hirzebruch surface $\IF_{n}$~.
These may be described as follows. We have homogeneous coordinates
$s,t,u,v,x,y,w$ acted on by three $\IC^\ast$'s $\lambda,\mu,\nu$ with weights
as in Table~\tabref{scaling}, and the manifold is realised as the vanishing
locus of a polynomial of the indicated degrees.
Higher members of the chains, \ie those corresponding to enhanced gauge
symmetry,
can be obtained by introducing A-D-E singularities in the generic fibres using
Tate's algorithm~
\REFS\rTate{J.~Tate, in: Modular Functions of One Variable IV, Lecture Notes
in Math.~vol.~476, Springer-Verlag, Berlin (1975) 33.}
\refsend.
The exact correspondence between the singularity type and
the coefficients of the Weierstrass equation has been worked out by
Bershadsky {\it et al.\/}~
\REFS\rBersh{M.~Bershadsky, K.~Intriligator, S.~Kachru, D.~R.~Morrison,
V.~Sadov and C.~Vafa, Nucl.Phys. B481 (1996) 215-252, hep-th/9605200.}
\refsend.
$$\vbox{\offinterlineskip\halign{
&\strut\vrule height 10pt depth 5pt #&\hfil\quad$#$\quad\vrule
&\hfil \quad$#$\quad&\hfil\qquad$#$\quad&\hfil\qquad$#$\quad
&\hfil\qquad$#$\quad&\hfil\qquad$#$\quad&\hfil\qquad$#$\quad
&\hfil\qquad$#$\quad\vrule&\quad$#$\quad\hfil\vrule\cr
\noalign{\hrule}
&&s&t&u&v&x&y&w&\hfil\hbox{degrees}\cr
\noalign{\hrule\vskip3pt\hrule}
&\l&1&1&\hidewidth{n}&0&\hidewidth{2n{+}4}&\hidewidth{3n{+}6}&0&6n+12\cr
&\m&0&0&1&1&4&6&0&12\cr
&\n&0&0&0&0&2&3&1&6\cr
\noalign{\hrule}
}}
$$
\nobreak\tablecaption{scaling}{The scaling weights of the elliptic
fibration over $\IF_{n}$.}
\bigskip
Consider the Weierstrass equation,
    $$ y^2 + a_1xy + a_3y = x^3 + a_2x^2 + a_4x + a_6 \eqlabel{Weier}$$
where the $a_i$'s are locally defined polynomial functions on the base. Denote
the affine coordinates of the $\IP_1$ base of the \cym\ and the $\IP_1$ base
of the elliptic $K3$ fibres by $z^\prime$ and $z$, respectively.
The singularities are then encoded in the degrees of vanishing of the
$a_i$'s. This has been worked out in detail in \cite\rBersh. We reproduce some
of their results in Table~\tabref{Singularities}.
For example, if we want an enhanced perturbative gauge group on the
heterotic side, then we locate the singularities at $z = 0$ and $z = \infty$,
corresponding to the first and second $E_8$'s, respectively.
It is easiest to explain this using a specific example. Consider the $n=0$
model, which is dual to the heterotic theory with 12 instantons in each
$E_8$.
This instanton configuration generically breaks the gauge group completely.
The Newton polyhedron\cite{\rBat,\rCOK} ~describing our \cym\ has 335
points, and the Hodge numbers are $(h_{11},h_{21}) = (3,243)$. If we now want
to unhiggs an $SU(2)$ subgroup of the first $E_8$, we proceed as follows:
\topinsert
\bigskip
 $$\vbox{\offinterlineskip\halign{\strut # height 10pt depth 5pt
&\quad$#$\quad\hfil\vrule
&\hfil\quad$#$\quad\hfil\vrule
&\hfil\quad$#$\quad\hfil\vrule
&\hfil\quad$#$\quad\hfil\vrule
&\hfil\quad$#$\quad\hfil\vrule
&\hfil\quad$#$\quad\hfil\vrule\cr
\noalign{\hrule}
\vrule&\hfil $Group$ & a_1 & a_2 & a_3 & a_4 & a_6 \cr
\noalign{\hrule\vskip3pt\hrule}
\omit{\vrule height2pt}&&&\cr
\vrule& SU(2)           & 0 & 0 & 1   & 1   & 2    \cr
\vrule& Sp(k)           & 0 & 0 & k   & k   & 2k   \cr
\vrule& SU(2k)          & 0 & 1 & k   & k   & 2k   \cr
\vrule& SU(2k+1)        & 0 & 1 & k   & k+1 & 2k+1 \cr
\vrule& G_2             & 1 & 1 & 2   & 2   & 3    \cr
\vrule& SO(4k+1)        & 1 & 1 & k   & k+1 & 2k   \cr
\vrule& SO(4k+2)        & 1 & 1 & k   & k+1 & 2k+1 \cr
\vrule& SO(4k+3)        & 1 & 1 & k+1 & k+1 & 2k+1 \cr
\vrule& SO(4k+4)^{\ast} & 1 & 1 & k+1 & k+1 & 2k+1 \cr
\vrule& F_4             & 1 & 2 & 2   & 3   & 4    \cr
\vrule& E_6             & 1 & 2 & 2   & 3   & 5    \cr
\vrule& E_7             & 1 & 2 & 3   & 3   & 5    \cr
\vrule& E_8             & 1 & 2 & 3   & 4   & 5    \cr
\omit{\vrule height2pt}&&&&&&\cr
\noalign{\hrule}
}}
$$
\nobreak\tablecaption{Singularities}{The relation between the degrees of
vanishing of terms in the Weierstrass Polynomial and the enhanced Gauge
Groups.}
\bigskip
\endinsert
The $SU(2)$ singularity corresponds to coefficients $a_i$ with
deg$(a_1,a_2,a_3,a_4,a_6) = (0,0,1,1,2)$,
where the numbers indicate the degrees of vanishing of the $a_i$'s near $z=0$.
We can introduce this singularity by noting that when the Weierstrass
equation~\eqref{Weier} is written in terms of the homogeneous coordinates,
the terms
labelled by $a_i$ contain a factor of $w^i$. Also, noting that $z = u/v$, we
see that if the degree of vanishing of $a_i$ near $z=0$ is $n$, then the
lowest power of $u$ in $a_i$ is also $n$. Thus the $SU(2)$ singularity is
obtained by discarding from the Newton polyhedron, the points with powers
$(n,i)$ of $u$ and $w$, respectively, with $n<(0,0,1,1,2)$ for
$i=(1,2,3,4,6)$. The Hodge numbers are now $(4,214)$.
The generalization to other gauge groups, as well as unhiggsing to subgroups
of the second $E_8$, is straightforward.
It is worth mentioning here that the new polyhedron does not describe a
singular manifold. Rather, it describes the
{\it smooth} \cym\ resulting from the resolution of the above singularity.
\subsection{\cys\ for the B and C series}
The \cys\ of the B and C chains were constructed in a similar way in
ref~\cite\rABCD. The \cys\ of the B chains were described as elliptic
fibrations over $\IF_{n}$, with the fibre the torus in
$\IP_{2}^{(1,1,2)}[4]$~. Similarly, the \cys\ of the C chains were described
as elliptic fibrations with the fibre the torus in $\IP_{2}[3]$. However, this
fibration structure was not actually used to construct the \cys\ in
ref~\cite\rABCD, but was only observed to hold in the examples studied. The
examples themselves were obtained by matching the expected hodge numbers with
those in a list of \cys .

We first attempted to construct higher members of these
chains by requiring the \cys\ to be elliptic fibrations as described above,
and then attempting to find  reflexive polyhedra satisfying these
conditions. However, we were not able to find many examples. In particular,
for the B series with $n=2$, although we found about 25 reflexive polyhedra,
we were only able to identify gauge groups $U(1)$ and $G\times U(1)$,
for $G = {SU(4), SO(8), E_6, E_7}$. We were not able
to identify the other reflexive polyhedra corresponding to elliptic \cys\
(with fibres a $\IP_2^{(1,1,2)}[4]$) with any heterotic model. It turns out,
however, that we can identify
more gauge groups if we consider not the torus in $\IP_2^{(1,1,2)}$ but rather
in a blowup of this space. The polyhedron of this torus, together with the
polyhedra of the A and C tori, are shown in fig\figref{3tori}.
Then it is possible to identify gauge groups
$G\times U(1)$, with
$$G = \{SU(1), SU(2), SU(3), SU(4), SO(8), SO(10), E_6, E_7\}~.$$
There were quite a few models which we could
not identify with gauge groups. Furthermore, some gauge groups were
``missing'', such as
$SU(5)\times U(1)$. We list the the Hodge numbers obtained with this torus
in Table~\tabref{allbref}
along with the corresponding gauge groups whenever they could be identified.
This table also lists hodge numbers computed using the torus of the B series
from Fig~\figref{3moretori} (refer to section 4.3), for comparison.

Similarly, for the C series with $n=2$, we only found the model with gauge
group $U(1)^2$. The only other reflexive polyhedron with fibre corresponding
to the torus in $\IP_2[3]$ had Hodge numbers corresponding to the B
model with gauge group $SU(4)\times U(1)$. Thus while the procedure for
obtaining models by constructing reflexive polyhedra produces a large number
of members of the B series, we were only able to construct the lowest member
of the C series.

These polyhedra were constructed by requiring that the polyhedra corresponding
to the $K3$ fibres project onto the polyhedron of the torus, a criterion which
was also used to construct the polyhedra of ref\cite\rCF. This
restriction is not actually necessary, and it is probable that the number of
reflexive polyhedra obtained may be increased by relaxing this condition,
though we have not investigated this in a systematic way. In the next section,
we present a different approach for constructing a rich class of
models, especially in the C series.

\figbox{\leavevmode\epsfxsize=5.8truein\hbox{%
\hskip.2in\epsfbox{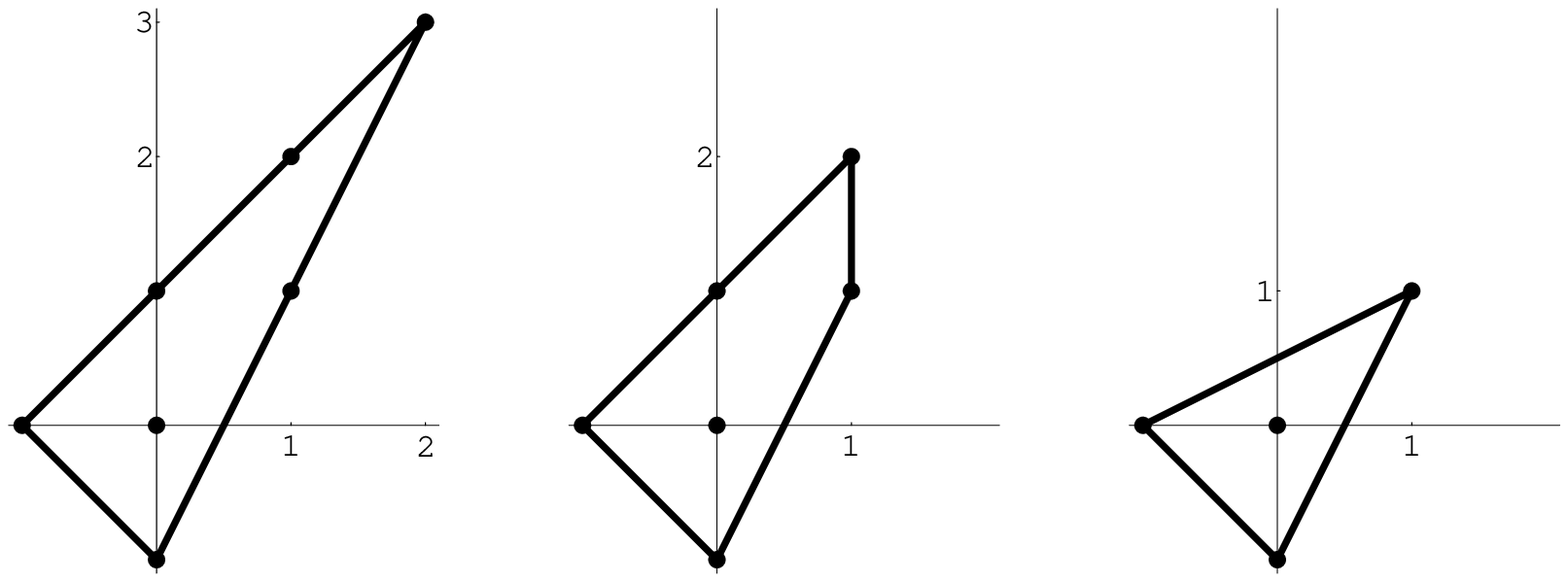}}}
{\figlabel{3tori}}{Polyhedra of the tori for the A, B and C
series.}
\place{1.8}{2.95}{$pt_1$}
\place{1.2}{2.50}{$pt_2$}
\place{0.7}{2.05}{$pt_4$}
\place{0.2}{1.55}{$pt_6$}
\place{1.45}{1.75}{$pt_3$}
\place{0.7}{0.8}{$pt_7$}
\place{0.7}{1.5}{$pt_5$}
\subsection{Conifold Transitions}
It was pointed out already in~\cite\rABCD ~that the A,B,C and D chains are
connected to each other by conifold transitions. This was shown explicitly for
the lowest members of the A and B chains for the case $n=4$. It was argued that
the chains should all be related by such transitions.
This now provides a useful way to actually construct the higher members
(those corresponding to enhanced gauge groups) of the chains. Since the A
chains are well understood, if we can induce an extremal transition to the
B,C and D chains by means of singularities (by restricting the form of
the monomials that define the manifolds in the A chains), we will then have
constructed candidate duals for the heterotic B, C and D chains. By comparing
the Hodge numbers of the \cys\ thus obtained with the data on the heterotic
side, we will be able to verify the conjectured duality.

In terms of the homogeneous coordinates $s,t,u,v,x,y,w$ defined above, the
transition from the A model to the B model is effected by adjusting the complex
structure of the manifold so that the equation may be written (with G and F
polynomials in $s,t,u,v,x,y,w$) as
$$ xG - yF = 0~. $$
Noting that the relative scalings of $x,y,w$ are $2{:}3{:}1$ respectively, and
that the manifold condition implies that the powers of $x,y$ and $w$ add up to
$6$, we see that the B model is obtained by restricting the Newton polyhedron
of the A model to points whose last coordinate (\ie the power of $w$) is less
than or equal to $4$.

Similarly, it turns out that it is possible to induce an extremal transition
that takes the A model to the C model. This happens when we adjust the complex
structure of the manifold so that the equation reads
$$ x^2G - yF = 0~. $$
This is equivalent to restricting the Newton
Polyhedron of the A model to points whose last coordinate is less than or
equal to $3$.

As with the \cys\ corresponding to enhanced gauge groups, it is important to
note that the Newton polyhedron obtained by the above method describes the
{\it smooth} manifold obtained by resolving the singularities introduced into
the original manifold .

The Hodge numbers of the \cys\ obtained by the above construction precisely
match the values for the spaces given in \cite\rABCD. That is, for
$n=0,\ldots 8$, the
lowest members of the A series map to the lowest members of the B series given
in \cite\rABCD , while for
$n>8$, the polyhedra obtained are non-reflexive, hence do not describe a
manifold. Similarly, for
$n=0\ldots 6$, the lowest members of the A chains map to the lowest members of
the C chains (except for $n=4$ --- see the second footnote to
Table~\tabref{Chodgenos}), while for $n>6$, the polyhedra obtained are
non-reflexive$^1$.
\footnote\ {$^1$Actually, it is only the Hodge numbers that match. The
polyhedra obtained are very different. The manifolds
obtained by the above method are birationally equivalent~
\REFS\rK{S.~Katz, private communication.}
\refsend\
to those described in~
\cite\rABCD.}
\subsection{Reflexive Polyhedra}
We now describe the sequences of reflexive polyhedra corresponding to the
above models and discuss the qualitative differences between the polyhedra
for the A, B and C series constructed as in the previous section. (Note that
the polyhedra described below are the
polyhedra dual to the Newton polyhedra of the previous section. We
hope that this will not cause confusion.)

The polyhedra of the A series were described in
{}~\cite\rCF. Apart from two points, which for
each member of the $n$'th chain of the A series are $(-1, 0, 2, 3)$ and
$(1, n, 2, 3)$, the points of the polyhedron
lie in the plane $x_1 = 0$ forming the polyhedron, ${}^3\nabla$ of the $K3$.
For each member of a chain, the polyhedron of the $K3$ is again divided
into a top and a bottom by the  polyhedron
${}^2\nabla$ of the torus and we may write
$$
{}^3\nabla^{n,H} = {}^3\nabla_{\hbox{top}}^n\cup
{}^3\nabla_{\hbox{bot}}^H ~,$$
where ${}^3\nabla_{\hbox{top}}^n$ depends only on $n$ while
${}^3\nabla_{\hbox{bot}}^H$ depends only on the group $H$ that is
perturbatively un-Higgsed in the heterotic side.
In particular then, the dual polyhedron ${}^4\nabla^{n,SU(1)}$ of the
lowest member of each chain can be written as
$${}^4\nabla^{n,SU(1)} = {}^3\nabla^{n,SU(1)} \cup
\{ (-1,0,2,3),\,(1,n,2,3)\} ~. $$

We can describe the tops and bottoms of ${}^3\nabla^{n,H}$ quite simply. There
is an interesting symmetry between the tops and bottoms. The tops of the
polyhedra describe the terminal (unbroken) gauge groups, which are subgroups
of the first $E_8$ and are listed below. The bottoms describe the enhanced
gauge groups which are subgroups of the second $E_8$.
$$\vbox{\offinterlineskip\halign{
\strut # height 10pt depth 5pt
&\quad$#$\quad\hfil\vrule
&\quad$#$\quad\hfil\vrule
&\quad$#$\quad\hfil\vrule
&\quad$#$\quad\hfil\vrule\cr
\noalign{\hrule}
\vrule&\hfil n&\hfil $A Series$&\hfil $B Series$&\hfil $C Series$\cr
\noalign{\hrule\vskip3pt\hrule}
\vrule&0,\,1,\,2        & SU(1)& U(1)& U(1)^2 \cr
\vrule&3                & SU(3)& SU(3){\times} U(1)& SU(3){\times} U(1)^2\cr
\vrule&4                & SO(8)& SO(8){\times} U(1)& SO(10){\times} U(1)^2\cr
\vrule&5                & F_4  & E_6{\times} U(1) & E_6{\times} U(1)^2\cr
\vrule&6                & E_6  & E_6{\times} U(1)& E_6{\times} U(1)^2\cr
\vrule&7,\,8            & E_7  & E_7{\times} U(1)& ---\hfil\cr
\vrule&9,\,10,\,11,\,12 & E_8  & ---\hfil& ---\hfil\cr
\noalign{\hrule}
}}
$$
\nobreak\tablecaption{tops}{The terminal groups for the A, B and C series.
Note that the groups do not in all cases simply acquire extra $U(1)$ factors.}
\newpage
The bottoms ${}^3\nabla_{\hbox{bot}}^H ~$, may be specified by
giving the points that lie below the torus ${}^2\nabla$. We label the points of
the torus as indicated in figure~\figref{3moretori}.
Also, let $pt_r^{(j)}$ represent j points of the lattice directly below the
corresponding points of ${}^2\nabla$. Thus, $pt_1^{(2)}$ represents the points
$(0,-1,2,3)$ and $(0,-2,2,3)$. Then the bottoms for each group can be specified
by giving the set of $pt_r^{(j)} ~$'s corresponding to the gauge group, as
shown in Table~\tabref{bots}. One might naively expect that the top for a
given (terminal) gauge group consists of the same points as the bottom, except
placed above the torus instead of below it. While this is not true in general,
it turns out that we can modify the tops so that they are indeed symmetric
with the corresponding bottoms, without changing the hodge numbers of the
manifold. It turns out that the number of extra points added is accompanied by
a decrease in the number of non-toric deformations of the polyhedra, so that
the \cym\ is unchanged.

Having described the structure of the reflexive polyhedra in the A series,
it is now easy to describe the structure of the polyhedra in the B and C
series.

\figbox{\leavevmode\epsfxsize=5.8truein\hbox{%
\hskip.2in\epsfbox{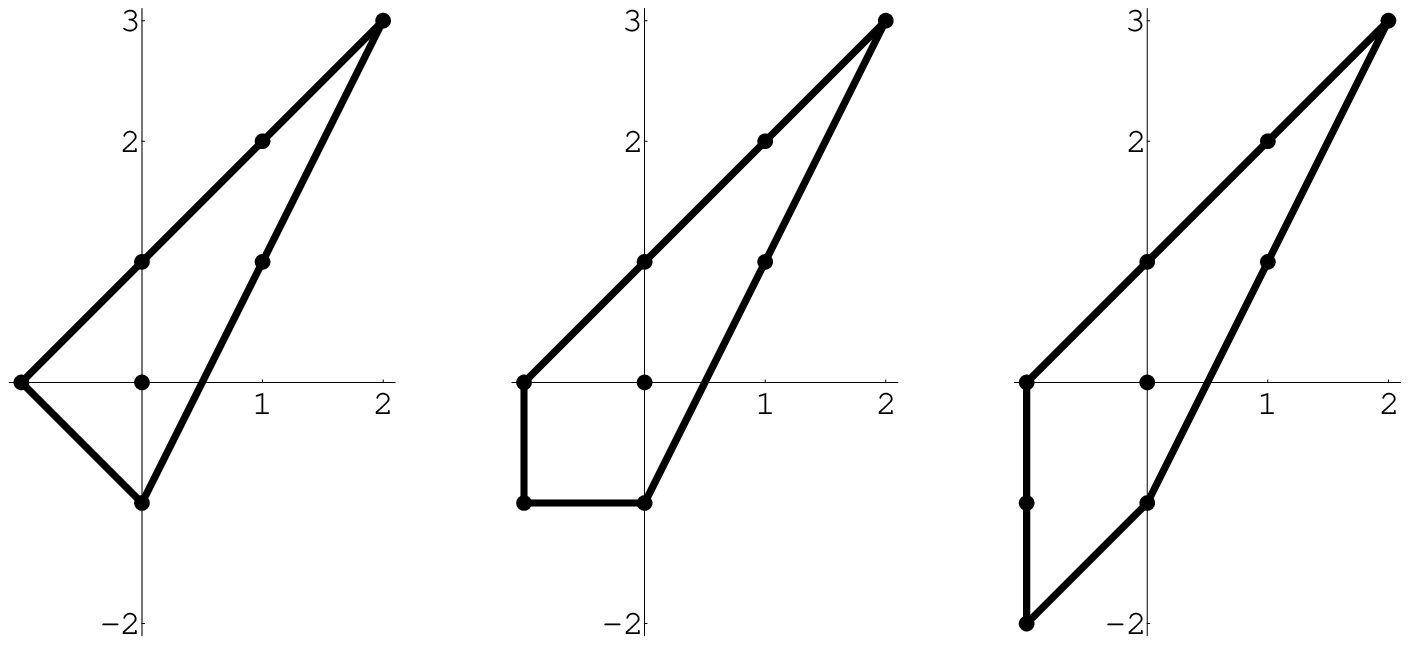}}}
{\figlabel{3moretori}}{The tori for the A series and its conifolds.}
\place{1.9}{3.45}{$pt_1$}
\place{1.2}{3.0}{$pt_2$}
\place{1.4}{2.3}{$pt_3$}
\place{0.6}{2.4}{$pt_4$}
\place{0.65}{2.0}{$pt_5$}
\place{0.15}{2.0}{$pt_6$}
\place{1.0}{1.45}{$pt_7$}
\place{2.15}{1.55}{$pt_8$}
\place{4.15}{1.55}{$pt_8$}
\place{4.15}{1.1}{$pt_9$}
Consider first the B series, constructed as described in section 4.3.
We have found that the general structure of the polyhedra of the B series
is similar to that of the A series. That is, there are always exactly two
4-dimensional points, which in our basis are (1,0,2,3) and (-1,0,2,3). Also,
the three dimensional polyhedron formed by the points with the first
coordinate equal to zero form a reflexive polyhedron, which we recognise as
encoding the $K3$ of the fibration. Furthermore, the points with first two
coordinates zero are always those shown in fig.~\figref{3moretori}.

$$\vbox{\offinterlineskip\halign{\strut # height 11pt depth 6pt
&\quad$#$\quad\hfil\vrule
&\quad$#$\quad\hfil\vrule\cr
\noalign{\hrule}
\vrule&\hfil $Group$&\hfil $Points in the Bottom$\cr
\noalign{\hrule\vskip3pt\hrule}
\omit{\vrule height13pt depth6pt}
& SU(2)&\{pt_1^{(1)}, pt_2^{(1)}\}             \cr
\vrule& SU(3)&\{pt_1^{(1)}, pt_2^{(1)}, pt_3^{(1)}\} \cr
\vrule& SU(4)&\{pt_1^{(1)}, pt_2^{(1)}, pt_3^{(1)}, pt_4^{(1)}\}\cr
\vrule& SU(5)&\{pt_1^{(1)}, pt_2^{(1)}, pt_3^{(1)}, pt_4^{(1)}, pt_5^{(1)}\}\cr
\vrule&SO(10)&\{pt_1^{(2)}, pt_2^{(2)}, pt_3^{(1)}, pt_4^{(2)}, pt_5^{(1)}\}\cr
\vrule& E_6& \{pt_1^{(3)}, pt_2^{(2)}, pt_3^{(2)}, pt_4^{(1)}, pt_5^{(1)}\}\cr
\vrule& E_7& \{pt_1^{(4)}, pt_2^{(3)}, pt_3^{(2)}, pt_4^{(2)}, pt_5^{(1)}\}\cr
\vrule& E_8& \{pt_1^{(6)}, pt_2^{(4)}, pt_3^{(3)}, pt_4^{(2)}, pt_5^{(1)}\}\cr
\vrule& SO(9)&\{pt_1^{(2)}, pt_2^{(2)}, pt_3^{(1)}, pt_4^{(1)}\}\cr
\vrule& F_4&\{pt_1^{(3)}, pt_2^{(2)}, pt_3^{(1)}, pt_4^{(1)}\}\cr
\noalign{\hrule}
}}
$$
\nobreak\tablecaption{bots}{The relation between the bottoms and the
enhanced gauge groups for the A series.}
\bigskip
While we find that the point labeled $pt_8$ is always present in the polyhedra
of the B series, this is not the only change in going from the A to the B
series. The polyhedra of the $K3$ fibres are also different --- extra points
are added to the tops, but the bottoms remain the same. This is because the
terminal groups change upon going from the A series to the B series. Just as
in the A series, we can modify the tops so that they become symmetric with the
bottoms of the same gauge groups. The terminal groups are listed in
Table~\tabref{tops}.

We have also studied \cys\ of the B series corresponding to enhanced gauge
groups. We find that, in general, the conifold transition described in
Sec.~3.2 maps a member of the A series corresponding to enhanced gauge group
$G$ to a member of the B series corresponding to enhanced gauge group
$G\times U(1)$. There are, however, some exceptions. For instance, the A model
with gauge group $F_4$ maps to the B model with gauge group $E_6\times U(1)$.
(The gauge
groups on the B series side are identified by their hodge numbers as predicted
by the heterotic calculations of Sec.~2). We list the gauge groups in the B
series obtained by conifolding the A series in Table~\tabref{ABgps}. Notice
that these are consistent with the terminal gauge groups listed in
Table~\tabref{tops}.

We list the tables of Hodge numbers of the B models in
Table~\tabref{Chodgenos} at the end of this section. We also present views of
the bottoms of the polyhedra corresponding to different enhanced gauge groups
in figures~A.1 and~A.2. We find that the Dynkin diagrams of the gauge groups
are visible in the polyhedra, as observed in~\cite\rCF\ for the A series.

\midinsert
\vskip10pt
$$\vbox{\offinterlineskip\halign{\strut # height 10pt depth 5pt
&\quad$#$\quad\hfil\vrule
&\quad$#$\quad\hfil\vrule
&\quad$#$\quad\hfil\vrule\cr
\noalign{\hrule}
\vrule&\hfil $A series groups$&\hfil $B series groups$&\hfil
             $C series groups$\cr
\noalign{\hrule\vskip3pt\hrule}
\vrule& SU(2) & SU(2)\times U(1) & SU(2)\times U(1)^2 \cr
\vrule& SU(3) & SU(3)\times U(1) &SU(3)\times U(1)^2 \cr
\vrule& SU(2k)& SU(2k)\times U(1)& SU(2k)\times U(1)^2 \cr
\vrule& SU(2k+1)& SU(2k+1)\times U(1)& SU(2k)\times U(1)^2\cr
\vrule& Sp(k)& Sp(k)\times U(1)& Sp(k)\times U(1)^2\cr
\vrule& G_2 & SO(7)\times U(1)& SO(10)\times U(1)^2\cr
\vrule& SO(7)& SO(7)\times U(1)& SO(10)\times U(1)^2\cr
\vrule& SO(8)& SO(8)\times U(1)& SO(10)\times U(1)^2\cr
\vrule& SO(9)& SO(10)\times U(1)& SO(10)\times U(1)^2\cr
\vrule& SO(10)& SO(10)\times U(1)& SO(10)\times U(1)^2\cr
\vrule& F_4 & E_6\times U(1)& E_6\times U(1)^{2} \, ^{\ast} \cr
\vrule& E_6 & E_6\times U(1)& E_6\times U(1)^{2} \, ^{\ast} \cr
\vrule& E_7 & E_7\times U(1) \, ^{\ast}& $(not reflexive)$ \cr
\noalign{\hrule}
}}
$$
\nobreak\tablecaption{ABgps}{The correspondence between gauge groups in the A,
B and C series. Asterisks indicate the presence of extra tensor multiplets.}
\bigskip
\endinsert

Consider next the C series, constructed as described in section 4.3.
We have found that the general structure of the polyhedra of the C series
is also similar to that of the A series. That is, there are always exactly two
4-dimensional points, which in our basis are (1,0,2,3) and (-1,0,2,3). Also,
the three dimensional polyhedron formed by the points with the first
coordinate equal to zero form a reflexive polyhedron, which we recognise as
encoding the $K3$ of the fibration. Furthermore, the points with first two
coordinates zero are always those shown in Fig.~\figref{3moretori}.

While we find that the points labeled $pt_8$ and $pt_9$ are always present in
the polyhedra of the C series, this is again not the only change in going from
the A to the C series. The polyhedra of the $K3$ fibres are also different ---
extra points are added to the tops, but the bottoms remain the same. Once
again, this is because the terminal gauge groups change upon going from the A
series to the C series. Furthermore, it is possible to modify the tops
without changing the hodge numbers, so that they are symmetric with the bottoms
of the same gauge groups. The terminal gauge groups are listed in Table~
\tabref{tops}.

We have also studied \cys\ of the C series corresponding to enhanced gauge
groups. We find that, in general, the extremal transition described in
Sec.~3.2 maps a member of the A series corresponding to enhanced gauge group
$G$ to a member of the C series corresponding to enhanced gauge group
$G\times U(1)^2$. There are again some exceptions. For instance, the A
model with gauge group $SU(4)$ maps to the C model with gauge group
$SU(5)\times U(1)^2$.
We list the gauge groups
in the C series obtained by extremal transitions of the A series in
Table~\tabref{ABgps}. The hodge numbers of the C series are listed in
Table~\tabref{Chodgenos}.
\midinsert
$$
\def\skip{\hskip2pt}
\vbox{\offinterlineskip\halign{
\strut # height 10pt depth 5pt
&\hskip5pt $#$ \skip \hfil\vrule
&\hskip10pt  \eightrm # \skip \hfil
&\skip  \eightrm # \skip \hfil
&\skip  \eightrm # \skip \hfil
&\skip  \eightrm # \skip \hfil
&\skip  \eightrm # \skip \hfil
&\skip  \eightrm # \skip \hfil
&\skip  \eightrm # \hskip5pt\hfil
&\skip  \eightrm # \hskip5pt\hfil\vrule
\cr
\noalign{\hrule}
\vrule&n&\hfil $SU(1)$&\hfil $SU(2)$&\hfil $SU(3)$&\hfil $SU(4)$&\hfil $SU(5)$
&\hfil $SO(10)$&\hfil $E_6$&\hfil $E_7$\cr
\noalign{\hrule\vskip3pt\hrule}
\vrule&0 &(148, 4)  &(133,  5)  &(122,  6)  &(115,  7) &(106,  8)  &(103,  9)
         &(98,  10) & (91, 19)\cr
\vrule&1 &(148, 4)  &(127,  5)  &(112,  6)  &(103,  7)
         &(92,  8)  &(88,  9) &(82,  10)&(74, 20)\cr
\vrule&2 &(148, 4)  &(121,  5)  &(102,  6)  &(91,  7)
         &(78,  8)  &(73,  9) &(66,  10)&(57, 21)\cr
\vrule&3 &(152, 6)  &(119,  7)  &(96,  8)  &(83,  9)
         &(68,  10)  &(62,  11) &(54,  12)&(44, 24)\cr
\vrule&4 &(164, 8)  &(125,  9)  &(98,  10)  &(83,  11)
         &(66,  12)  &(59,  13) &(50,  14)&(39, 27) \cr
\vrule&5 &(178, 10)  &(133,  11)  &(102,  12)  &(85,  13)
         &(66,  14)  &(58,  15) &(48,  16)&(36, 30) \cr
\vrule&6 &(194, 10)  &(143,  11) &(108,  12) &(89,  13)
         &(68,  14)  &(59,  15) &(48,  16)&(35, 31)\cr
\vrule&7 &(210, 12) &(153,  13) &(114,  14) &(93,  15)
         &(70,  16)  &(60,  17) &(48,  18)&(34, 34)\cr
\omit{\vrule height 11pt depth6pt}
&8       &(227, 11) &(164,  12) &(121,  13) &(98,  14)
         &(73,  15) &(62,  16) &(49,  17)&(34, 34)\cr
\noalign{\hrule\vskip3pt\hrule}
\vrule&0 &(101, 5)  &(90,  6)  &(85,  7)  &(78,  8)$^{\ast}$
         &(73,  9)&(70,  10)&(65, 17)&\cr
\vrule&1 &(101, 5)  &(86,  6)  &(79,  7)  &(70,  8)$^{\ast}$
         &(64,  9)&(60,  10) &(54, 18)&\cr
\vrule&2 &(101, 5)  &(82,  6)  &(73,  7)  &(62,  8)$^{\ast}$
         &(55,  9)&(50,  10) &(43, 19)&\cr
\vrule&3 &(103, 7)  &(80,  8)  &(69,  9)  &(56, 10 )$^{\ast}$
         &(48,  11)&(42,  12) &(34, 22)&\cr
\omit{\vrule height 11pt depth6pt}
&4{^{\dag}} &(110, 10)$^{\dag}$
         &(83,  11)$^{\dag}$  &(70,  12)$^{\dag}$  &(55,  13)$^{\ast}{}^{\dag}$
         &(46,  14)$^{\dag}$&(39,  15)$^{\dag}$ &(30, 26)$^{\dag}$&\cr
\vrule&5 &(120, 12)  &(89,  13)  &(74,  14)  &(57,  15)$^{\ast}$
         &(47,  16)&(39,  17) &(29, 29)&\cr
\omit{\vrule height 11pt depth6pt}
&6       &(131, 11) &(96,  12) &(79,  13) &(60,  14)$^{\ast}$
         &(49,  15)&(40,  16) &(29, 29)&\cr
\noalign{\hrule}
}}
$$
\nobreak\kern.75truein\lower0truept\hbox{\vbox{\hsize=5truein
    \noindent{\bf Table\hskip5truept\tablabel{Chodgenos}:} The Hodge numbers
$(h_{21},  h_{11})$ for the B and C series.}}
\endinsert
\pageinsert
$${\def\skip{\hskip4pt}
\vbox{\offinterlineskip\halign{\strut # height 9pt depth 5pt
&\skip  $#$ \hskip3pt\hfil\vrule
&\skip  $#$ \hskip3pt\hfil\vrule
&\skip  $#$ \skip\hfil\vrule\cr
\noalign{\hrule}
\omit{\vrule height11pt depth7pt}
&\hfil (h_{21}, h_{11})&\hfil(h_{21}, h_{11}) &\hfil $Gauge Groups$\cr
\noalign{\hrule\vskip3pt\hrule}
\omit{\vrule height11pt depth5pt}
& (100, 4)&         &       \cr
\vrule& (118, 4)&         &       \cr
\vrule& (148, 4)&  (148, 4)& SU(1){\times} U(1)\cr
\vrule& (75, 5)&          &       \cr
\vrule& (89, 5)& (89, 5)  &       \cr
\vrule& (93, 5)&          & SU(2)_b{\times} U(1)\cr
\vrule&        & (101,5)  &\cr
\vrule& (111, 5)&         & SU(2)_c{\times} U(1)\cr
\vrule& (121, 5)&(121, 5) & SU(2){\times} U(1)\cr
\vrule&         &(70, 6)  &\cr
\vrule& (72, 6)&          & \cr
\vrule& (78, 6)&          &\cr
\vrule& (82, 6)& (82, 6)  &\cr
\vrule& (88, 6)&          &\cr
\vrule& (90, 6)&          & SU(3)_b{\times} U(1)\cr
\vrule&        & (98, 6)  & Sp(2){\times} U(1)\cr
\vrule& (102, 6)& (102, 6)& SU(3){\times} U(1)\cr
\vrule& (65, 7)&          &\cr
\vrule& (67, 7)& (67, 7)  &\cr
\vrule& (69, 7)&          &\cr
\vrule& (75, 7)&          &\cr
\vrule&        & (79, 7)  & Sp(3){\times} U(1)\cr
\vrule&        & (81, 7)  &\cr
\vrule& (85, 7)&          & SU(4)_b{\times} U(1)\cr
\vrule& (91, 7)& (91, 7)  & SU(4){\times} U(1)\cr
\vrule& (64, 8)& (64, 8)  & Sp(4){\times} U(1)\cr
\noalign{\hrule}
}}
\hskip20pt
\vbox{\offinterlineskip\halign{\strut # height 9pt depth 5pt
&\skip  $#$ \hskip3pt\hfil\vrule
&\skip  $#$ \hskip3pt\hfil\vrule
&\skip  $#$ \skip\hfil\vrule\cr
\noalign{\hrule}
\omit{\vrule height11pt depth7pt}
&\hfil (h_{21}, h_{11})&\hfil (h_{21}, h_{11})&\hfil $Gauge Groups$\cr
\noalign{\hrule\vskip3pt\hrule}
\omit{\vrule height11pt depth5pt}
& (66, 8)& (66, 8)  &\cr
\vrule& (70, 8)&          &\cr
\vrule&        & (72, 8)  &\cr
\vrule&        & (78, 8)  & SU(5){\times} U(1)\cr
\vrule& (80, 8)&          & SO(8){\times} U(1)\cr
\vrule&        & (57, 9)  &\cr
\vrule&        & (61, 9)  &\cr
\vrule& (63, 9)& (63, 9)  &\cr
\vrule& (65, 9)&          &\cr
\vrule&        & (69, 9)  &\cr
\vrule& (73, 9)& (73, 9)  & SO(10){\times} U(1)\cr
\vrule&        & (56, 10)  &\cr
\vrule&        & (58, 10)  &\cr
\vrule&        & (64, 10)  &\cr
\vrule& (66, 10)&(66, 10)  & E_6{\times} U(1)\cr
\vrule&        & (55, 11)  &\cr
\vrule& (60, 12)&         &\cr
\vrule& (62, 12)&         &\cr
\vrule&        & (49, 13)  &\cr
\vrule&        & (54, 14)  &\cr
\vrule& (59, 15)&         &\cr
\vrule& (61, 15)&         &\cr
\vrule&         & (53, 17)&\cr
\vrule& (58, 18)& (58, 18)& E_6{\times} U(1) + 8 \hbox{T}\cr
\vrule& (57,21)&  (57, 21)& E_7{\times} U(1) + 10\hbox{T}\cr
\vrule&&&\cr
\noalign{\hrule}
}}
}$$
\nobreak\tablecaption{allbref}{Hodge numbers for the B series with $n=2$
obtained by two different means. The first column gives the values obtained
using the torus of the B series from fig~\figref{3tori} and the second gives
the values using the torus of the B series from fig~\figref{3moretori}. The
third column lists the gauge groups, where known. Blank entries in either of
the first two columns indicate missing hodge numbers, while in the third
column they indicate unidentified gauge groups. In the last two entries, the
presence of extra tensor multiplets is noted.}
\bigskip
\endinsert
Note that we do not observe the spaces marked with a~$^{\ast}$ by
the methods described in Sec.~3.2.
The Hodge numbers listed here have not been observed by us.
Rather, they were calculated using the heterotic data. If these spaces were
indeed to exist, they would have these Hodge numbers. Furthermore, the Hodge
numbers for the $n=4$ case (marked with a~$^{\dag}$) are those
obtained by extremal transitions. However, this method yields terminal
gauge group $SO(10)\times U(1)^2$ (see Table~\tabref{ABgps}), but the
terminal gauge group obtained in~\cite\rABCD ~is $SO(8)\times U(1)^2$. There is
thus a discrepancy of one for each of the $(h_{21}, h_{11})$ in this chain.
For instance, the lowest member of the chain should have Hodge numbers
$(111,9)$ rather than $(110,10)$ as listed here.

Views of the polyhedra corresponding to different
enhanced gauge groups are presented in figures~A.3 and~A.4. Once again, we
find that the Dynkin diagrams of the gauge groups are visible in the
polyhedra, as observed in~\cite\rCF\ for the A series.

\subsection{Nonperturbative Vacua --- Tensor Multiplets}
The previous sections described F-theory vacua dual to perturbative heterotic
vacua. It is also straightforward to describe vacua corresponding to extra
tensor multiplets, which are nonperturbative vacua on the heterotic side.
Briefly, the extra tensor multiplets in the A series are obtained by blowing
up the base $\IF_{n}$ of the \cym\ \cite\rMV.
This is achieved by adding extra lines of weights to the weight systems that
describe the Newton polyhedron~
\REFS\rBS{P.~Candelas, E.~Perevalov and G.~Rajesh, hep-th/9606133.}
\refsend.
This results in extra
points being added to the fan of the $\IF_{n}$ in the dual polyhedron in
such a way that the change in the $h_{11}$ of the $\IF_{n}$ (which can be
computed easily --- see ref.~\REFS\rFul{W.~Fulton, Introduction to Toric
Varieties,\\
Princeton University Press, Princeton, 1993.}
\refsend\ ) is precisely the required number of extra tensor multiplets.

Combining this method with the conifold transitions of Sec.~3.2, we can now
construct B and C series vacua with extra tensor multiplets. We simply
introduce a given number of extra tensor multiplets on the A series and then
apply the conifold transitions to obtain the B and C series vacua with the
same number of extra tensor multiplets.
We have found that the Hodge numbers of these \cys\ follow a simple pattern.
For the B series, the
creation of an extra tensor multiplet increases $h_{11}$ by 1, while reducing
$h_{21}$ by 17, which is one less than the dual Coxeter number of $E_7$. This
agrees with the results of~\cite\rABCD. Of course, creation of an extra tensor
multiplet removes an $SU(2)$ instanton, so that the terminal gauge group may
change, thus, the change in the Hodge numbers may not be simply given by
$({\Delta}h_{21}, {\Delta}h_{11}) = (-17, 1)$. The correction, when it exists,
may be obtained by working out the new terminal groups using the data listed in
Table~\tabref{Bhiggs}.

Additional subtleties arise when there would be less than three $SU(2)$
instantons
in any $E_8$. These have been discussed in detail in \cite\rABCD. The result is
that these situations cannot be realised. Where we would expect two $SU(2)$
instantons to be present, it turns out that one of these is replaced
by a $U(1)$ instanton, and the other by a tensor multiplet, leaving an
unbroken $E_7$ gauge group. Also, where we would expect one $SU(2)$ instanton,
it is replaced by a $U(1)$ instanton, again leaving an unbroken $E_7$ gauge
group. The Hodge numbers can be worked out easily from the heterotic
side by the methods discussed here and in \cite\rABCD\ and we have indeed found
that the \cys\ constructed as above have precisely these Hodge numbers. For
example starting with the $n=2$ model, which has $SU(2)$ instanton numbers
$(11, 7)$ in the two $E_8$'s respectively, removal of 9 or 10 instantons from
the first $E_8$ yields Hodge numbers (57, 21) --- the \cys\ are actually the
same because creation of 9 tensor multiplets results in the creation of a
tenth one, by the above considerations. Furthermore, it is not possible to
remove all the $SU(2)$ instantons from any $E_8$ since this yields a member of
the A series \cite\rABCD . In fact, what happens is that introduction of the
conifold singularity destroys the condition of vanishing first Chern class.

In Table~\tabref{Btm} we list the Hodge numbers for the \cys\ corresponding to
extra tensor multiplets in the $n=2$ model. These tensor multiplets are
obtained by removing $SU(2)$ instantons from the first $E_8$.
\midinsert
$$
\def\skip{\hskip1pt}
\vbox{\offinterlineskip\halign{
\strut # height 10pt depth 5pt
&\hfil\hskip3pt $#$ \skip \hfil\vrule
&\hskip1pt  \eightrm # \skip \hfil
&\skip  \eightrm # \skip \hfil
&\skip  \eightrm # \skip \hfil
&\skip  \eightrm # \skip \hfil
&\skip  \eightrm # \skip \hfil
&\skip  \eightrm # \skip \hfil
&\skip  \eightrm # \skip \hfil
&\skip  \eightrm # \skip \hfil
&\skip  \eightrm # \skip \hfil
&\skip  \eightrm # \skip \hfil\vrule\cr
\noalign{\hrule}
\vrule&$Tensors$&\hfil 0&\hfil 1&\hfil 2&\hfil 3&\hfil 4&\hfil 5&\hfil 6&
\hfil 7&
\hfil 8&\hfil 9, 10\cr
\noalign{\hrule\vskip3pt\hrule}
\vrule
&$B series$ &(148, 4)&(131, 5)&(114, 6)&(97, 7)
&(80, 8)&(67, 11)&(62, 14)&(59, 17)&(58, 18)&(57, 21)\cr
\noalign{\hrule}
\vrule
&$C series$ &(101, 5)&(90, 6)&(79, 7)&(68, 8)
&(57, 9)&(48, 12)&(44, 16)&(43, 19)&(43, 19)&\hskip10pt ---\cr
\noalign{\hrule}
}}
$$
\nobreak\tablecaption{Btm}{The Hodge numbers for the $n=2$ B and C models with
extra tensor multiplets.}
\endinsert
\smallskip
For the C series,
creation of an extra tensor multiplet increases $h_{11}$ by 1, while reducing
$h_{21}$ by 11, which is one less than the dual Coxeter number of $E_6$. This
agrees with the results of \cite\rABCD. Of course, creation of an extra tensor
multiplet removes an $SU(2)$ instanton, so that the terminal gauge group may
change, thus, the change in the Hodge numbers may not simply be given by
$({\Delta}h_{21}, {\Delta}h_{11}) = (-11, 1)$. Again, the correction, if any,
may be obtained by working out the new terminal groups using the data listed in
Table~\tabref{Bhiggs}.

Subtleties again arise when there would be less than three $SU(2)$ instantons
in any $E_8$. These situations cannot be realised. Where we would expect two
$SU(2)$ instantons to be present, it turns out that one of these is replaced
by a $U(1)$ instanton, and the other by a tensor multiplet, leaving an
unbroken $E_6$ gauge group. Also, where we would expect one $SU(2)$ instanton,
it is replaced by a $U(1)$ instanton, again leaving an unbroken $E_6$ gauge
group. The Hodge numbers can be worked out easily from the heterotic
side by the methods discussed here and in \cite\rABCD ~and we have indeed found
that the \cys\ constructed as above have precisely the expected Hodge numbers.
For example starting with the $n=2$ model, which has $SU(2)$ instanton numbers
$(9, 5)$ in the two $E_8$'s respectively, removal of 7 or 8 instantons from
the first $E_8$ yields Hodge numbers (43, 19) --- the \cys\ are actually the
same because creation of 7 tensor multiplets results in the creation of
another one. Once again, it is not possible to
remove all the $SU(2)$ instantons from any $E_8$ since we would then fall back
on the B series \cite\rABCD\ --- it is not possible to take the conifold of
this model without destroying the condition of vanishing first Chern class.

In Table~\tabref{Btm} we list the Hodge numbers for the \cys\ corresponding to
extra tensor multiplets in the $n=2$ C model. These tensor multiplets are
obtained by removing $SU(2)$ instantons from the first $E_8$.
\section{fin}{Discussion}
In this article we have studied F-theory duals of heterotic $E_8\times E_8$
compactifications on $K3$ with non-semisimple backgrounds of the type $H\times
U(1)$ or $H\times U(1)^2$ in each $E_8$, which were originally studied
in~\cite\rABCD , and were called B and C type models respectively. Our paper
extends the previous results by using toric geometry to construct many more
examples of B and C models corresponding to enhanced gauge symmetry, as well
as models with extra tensor multiplets. This description
also has the interesting consequence that the Dynkin diagrams of the gauge
groups are visible in the polyhedra, as observed in~\cite\rCF\ for the A
series.
We find that the \cys\ corresponding to such vacua are most
easily constructed by applying conifold-type extremal transitions to the A
models
of~\cite\rCF, and that construction of the B and C models can be systematised
easily.
\vskip5pt
\noindent {\bf Acknowledgements}
\vskip5pt
\noindent We wish to thank A.~Font for explaining the methods
discussed in section 3 of this paper.
This work was supported in part by the Robert Welch Foundation and NSF grant
PHY-9511632.
\chapno=-1
\section{appendix}{Appendix: Figures}
We append below figures of some of the bottoms of the polyhedra of the B and C
series. In the electronic version of this article the figures are in color.
The color coding is as in ref~\cite\rCF.
\newpage
\figbox{\leavevmode\epsfysize=7truein\hbox{%
\hskip1.2in\epsfbox{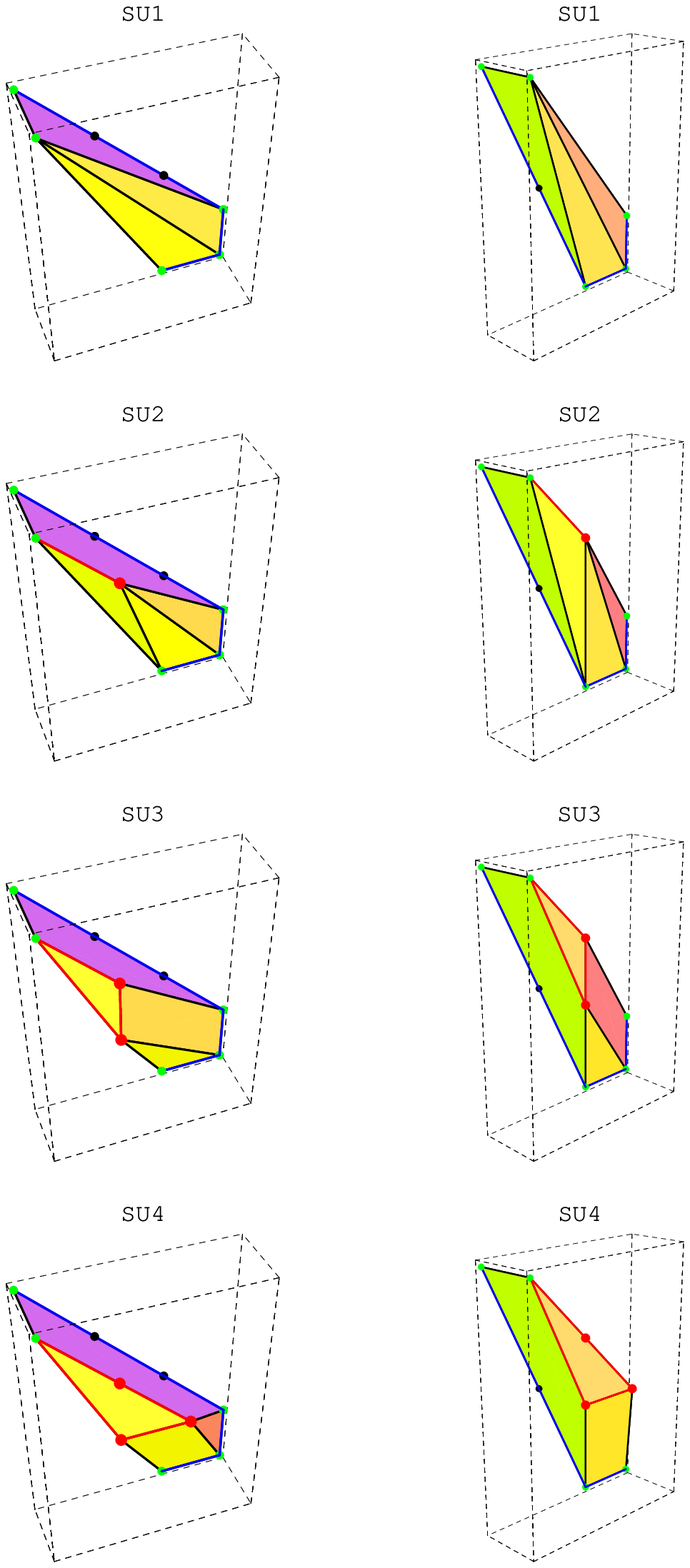}}}
{\figlabel{figca1}}{Two views of some of the polyhedra for the B series with
$n=2$.}
\newpage
{\figbox{\leavevmode\epsfysize=7.5truein\hbox{%
\hskip.75in\epsfbox{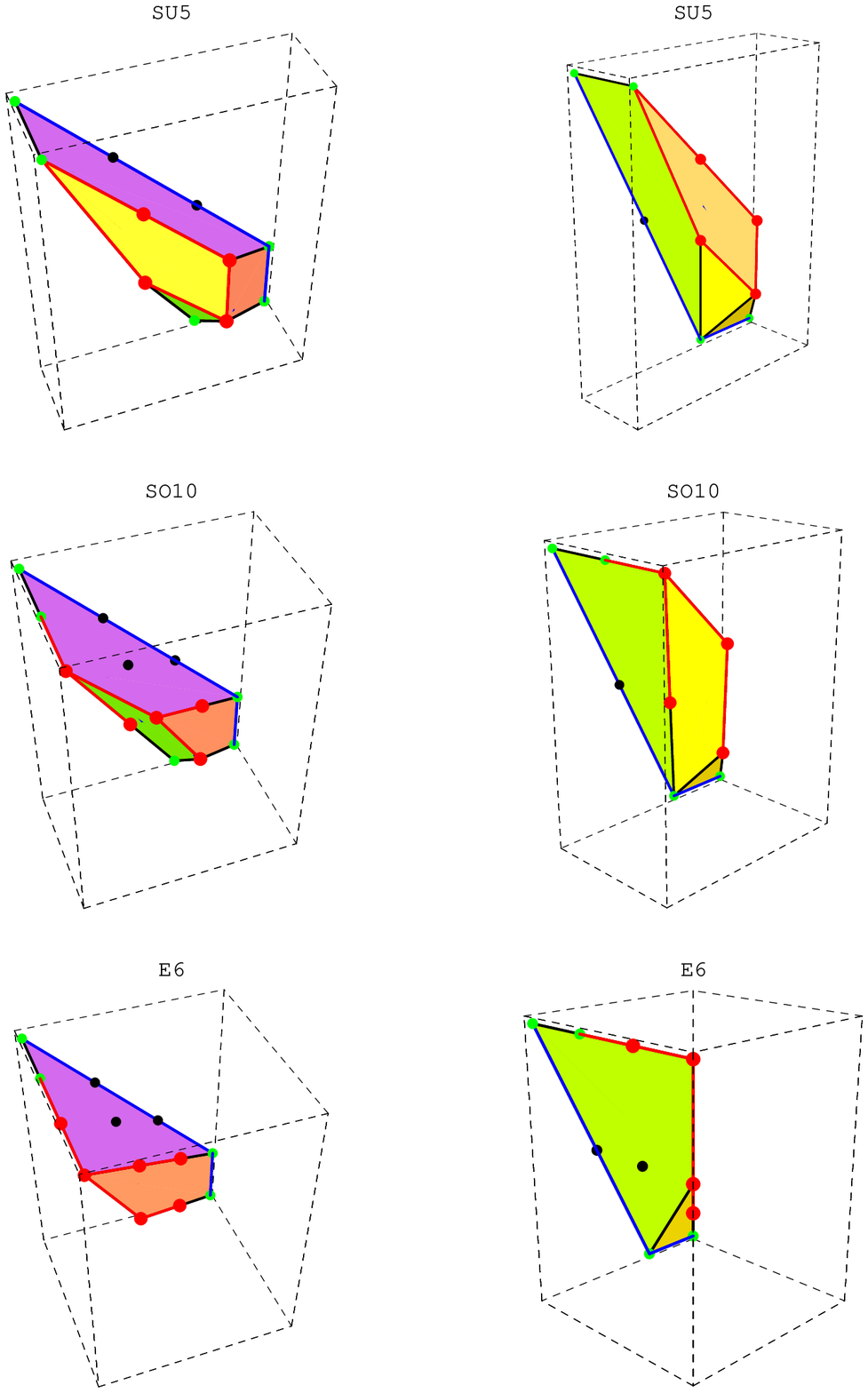}}}
{\figlabel{figca2}}{Two views of some of the polyhedra for the B series with
$n=2$.}
\newpage}
{\figbox{\leavevmode\epsfysize=7.5truein\hbox{%
\hskip.75in\epsfbox{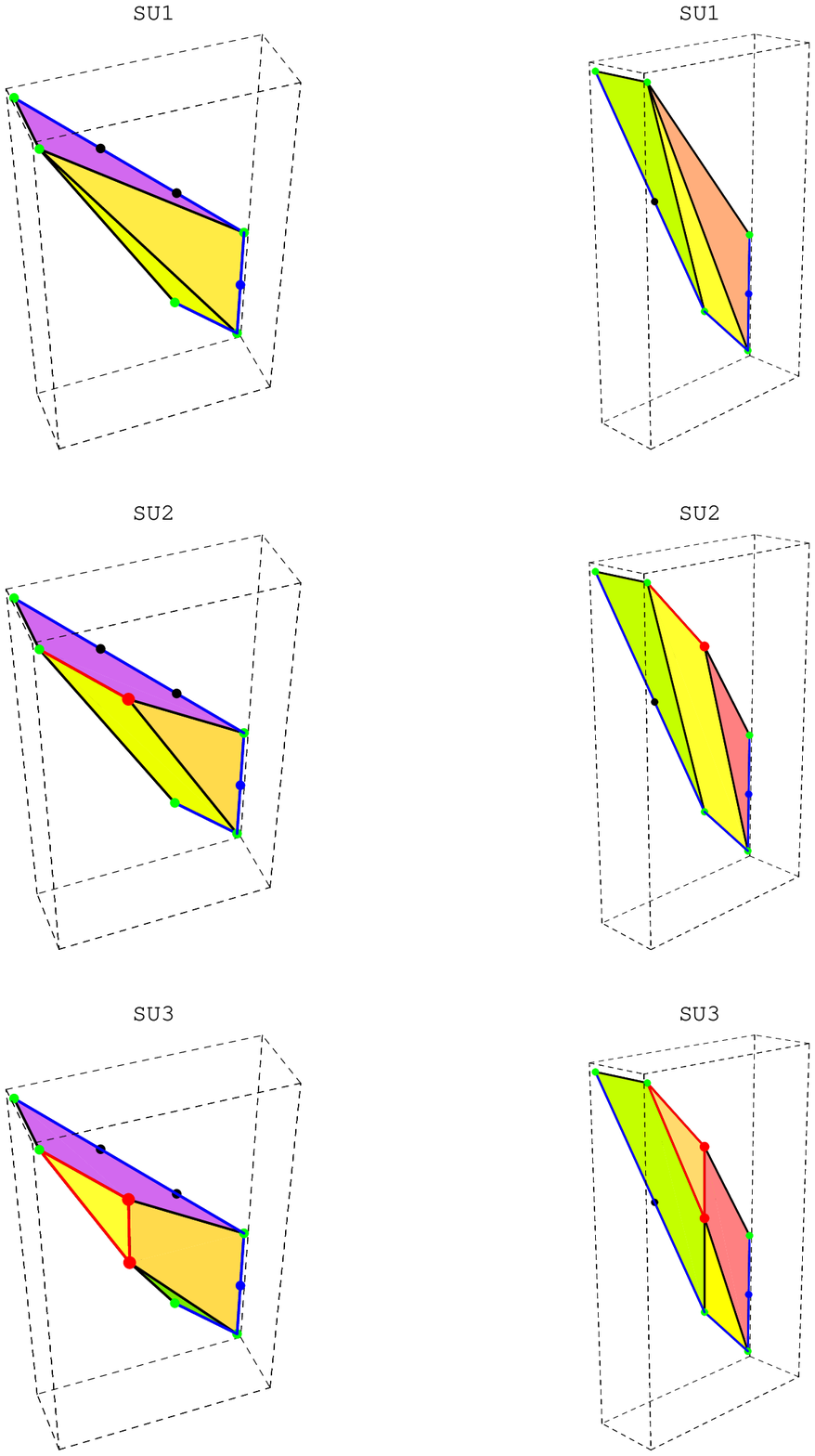}}}
{\figlabel{figcca1}}{Two views of some of the polyhedra for the C series with
$n=2$.}
\newpage}
{\figbox{\leavevmode\epsfysize=5truein\hbox{%
\hskip.75in\epsfbox{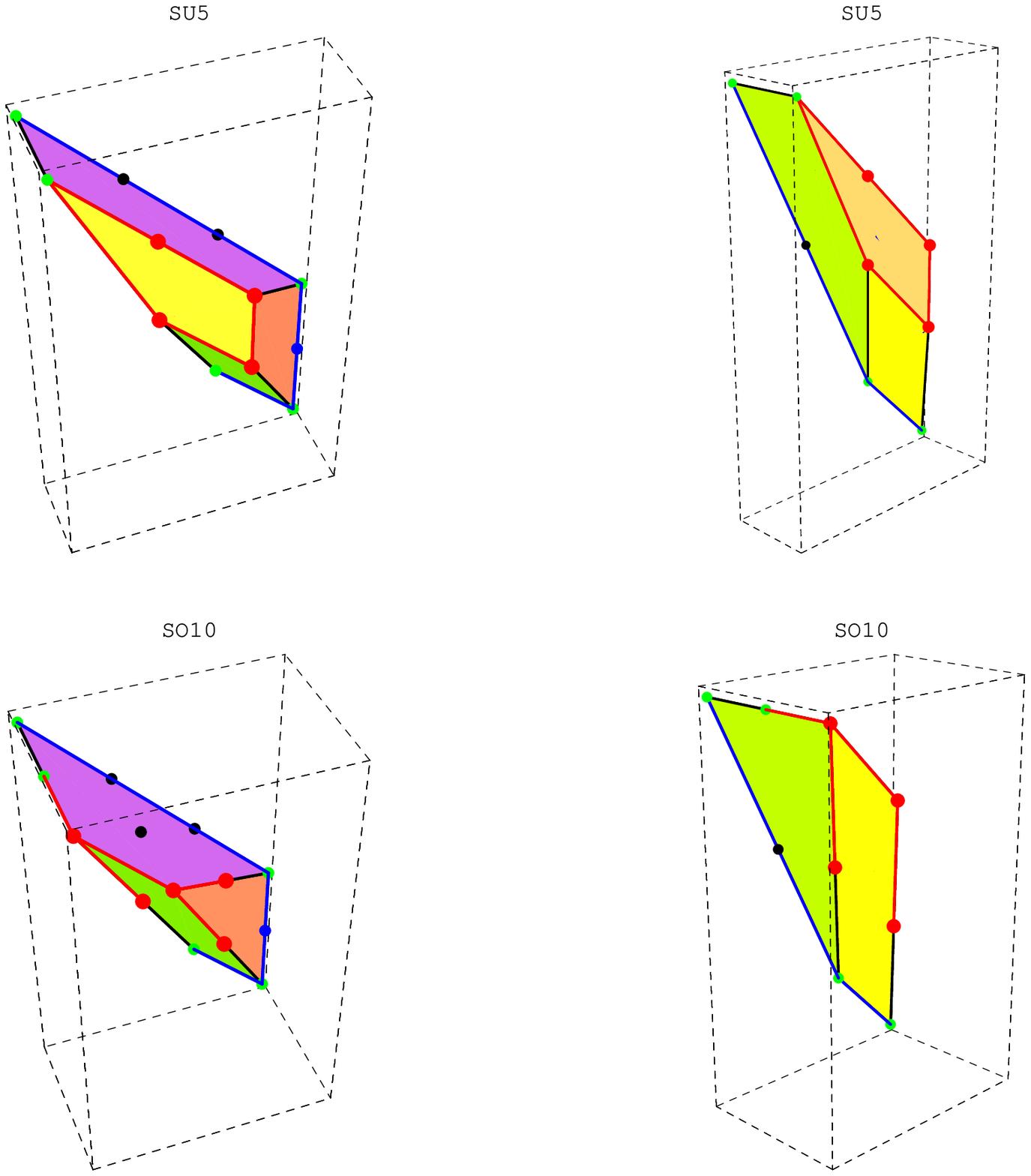}}}
{\figlabel{figcca2}}{Two views of some of the polyhedra for the C series with
$n=2$.}
\newpage}

\immediate\closeout\referencewrite\referenceopenfalse
\line{\bf\hfil References\hfil}\bigskip\parindent=0pt\input referenc.texauxil

\bye